\documentclass[11pt]{article}

\usepackage{epsfig,graphics}
\usepackage{amssymb,amsmath,amsthm,bm}
\usepackage{algorithm}
\usepackage{algorithmic}
\usepackage{subfigure}    
\usepackage{color}

\newtheorem{theo}{Theorem}

\newtheorem{mdef}{Definition}

\newtheorem{assum}{Assumption}

\def\ZZ{{\mathbb Z}}
\def\RR{{\mathbb R}}

\def\ga{\alpha}

\def\gth{\theta}
\def\go{\omega}

\def\gb{\beta}
\def\gd{\delta}

\def\gs{\sigma}
\def\gl{\lambda}
\def\wt{\widetilde}
\def\n{\noindent}

\def\gp{\prime}

\def\gt{\triangle}

\def\b0{{\bf 0}}
\def\1{{\bf 1}}

\def\cG{\mathcal G}
\def\cH{{\mathcal H}}
\def\cD{{\mathcal D}}

\def\vep{\varepsilon}

\def\wh{\widehat}

\def\wt{\widetilde}

\usepackage{scalerel,stackengine}
\stackMath
\newcommand\reallywidecheck[1]{%
\savestack{\tmpbox}{\stretchto{%
  \scaleto{%
    \scalerel*[\widthof{\ensuremath{#1}}]{\kern-.6pt\bigwedge\kern-.6pt}%
    {\rule[-\textheight/2]{1ex}{\textheight}}
  }{\textheight}%
}{0.5ex}}%
\stackon[1pt]{#1}{\scalebox{-1}{\tmpbox}}%
}
\def\wc{\reallywidecheck}

\begin{document}

\title{Direct Signal Separation Via Extraction of \\ Local Frequencies	with Adaptive \\ Time-Varying Parameters
}

\author{Lin Li, Charles K. Chui, and Qingtang Jiang\thanks{L. Li is with School of Electronic Engineering, Xidian University, Xi\rq{}an, China;
C.K. Chui lives in Menlo Park and is affiliated with the Department of Statistics, Stanford University, Stanford, CA, USA;
 and Q. Jiang is with Department of Math \& Statistics, University of Missouri-St. Louis, St. Louis,  MO, USA.
This work is partially supported by the National Natural Science Foundation of China under Grant \#62071349, the U.S. Army Research Office, under ARO Grant \#W911NF2110218, and the Simons Foundation, under Grant \#353185.}}
\date{}


\maketitle

\begin{abstract}

Real-world phenomena that can be formulated as signals are often affected by a number of factors and appear as multi-component modes. 
To understand and process such phenomena, ``divide-and-conquer" is probably the most common strategy to address the problem. In other words, the captured signal is decomposed into signal components for each individual component to be processed.
Unfortunately, for signals that are superimposition of non-stationary
amplitude-frequency modulated (AM-FM) components, the ``divide-and-conquer" strategy is bound to fail, since there is no way to be sure that the decomposed components take on the AM-FM formulations which are necessary for the extraction of their instantaneous frequencies (IFs) and amplitudes (IAs). 
In this paper, we propose an adaptive signal separation operation (ASSO) for effective and accurate separation of a single-channel blind-source multi-component signal, via 
introducing  a time-varying parameter that adapts locally to IFs and using linear chirp (linear frequency modulation) signals to approximate components at each time instant. We derive  more accurate component recovery formulae based on the linear chirp signal local approximation. In addition, a recovery scheme, together with a ridge detection method,  is also proposed to extract the signal components one by one, and the time-varying parameter is updated for each component. The proposed method is suitable for engineering implementation, being capable of separating complicated signals into their components or sub-signals and reconstructing the signal trend directly. Numerical experiments on synthetic and real-world signals are presented to demonstrate our improvement over the previous attempts.
\end{abstract}



\section{Introduction}

For communication, audio, and other applications, a single-channel multicomponent signal $x(t)$ is usually represented as a superimposition of Fourier-like oscillatory amplitude-frequency modulated (AM-FM) components, called the AM-FM model \cite{Flandrin10,AM_FM_Francesco07}. 
Motivated by the empirical mode decomposition (EMD) \cite{Huang98}, we model a multicomponent signal (or in terms of a time series) by
\begin{equation}
\label{EMD_model}
x(t)=\sum_{k=1}^K x_k(t) + A_0(t) =F(t)+A_0(t),
\end{equation}
where
\begin{equation}
\label{AM-FM}
F(t)=\sum_{k=1}^K x_k(t)=\sum_{k=1}^K A_k(t) \cos \left(2\pi \phi_k(t)\right).
\end{equation}
Here $A_0(t)$, called the trend of $x(t)$, is (at most) minimally oscillatory. The function $F(t)$ in \eqref{AM-FM} with $0<\mu \le A_k(t) \le M$, $\phi_k'(t)  \ge 0$ and $\phi_k'(t) > 0 $ almost everywhere (a. e.), $\phi_k'(t)>\phi_{k-1}'(t)$, and $A_k(t)$, $\phi_k'(t)$ varying more slowly than $\phi_k(t)$ (see \cite{Chui_M15}, Eq.(1.3)), is called the AF-FM model, where  $A_k(t)$ are called the instantaneous amplitudes (IAs)  and $\phi'_k(t)$ the instantaneous frequencies (IFs), which can be used to describe the underlying dynamics. In the mathematical literature,
$F(t)$ is also called the ``adaptive harmonic model (AHM)"  (see for example \cite{Chui_M15, Daub_Lu_Wu11, Wu_thesis}). Note that it is not necessary for the components $x_k$ to be defined on all of $\RR$. They may be supported on some finite or infinite sub-intervals of $\RR$ as considered in \cite{Daub_Lu_Wu11}.

The present paper is devoted to resolving the inverse problem of recovering the non-stationary signal components $x_k(t)$, $k=1, \cdots, K$ and trend $A_0(t)$ from the ``blind-source" data of the composite signal $x(t)$ governed by the model \eqref{EMD_model} and \eqref{AM-FM}, first by extracting the IFs $\phi_k'(t), k=1, \cdots, K$. For the stationary setting  (i.e. for $\phi_k'(t)$ being positive constants for all $t\in \RR$) with trend $A_0(t)=0$ and positive constant amplitudes $A_k$, the solution of this problem already appeared in the classic work \cite{De Prony} of De Prony (called Prony's method), where the number $K$ of components is assumed to be known. Significant improvements of \cite{De Prony} to allow unknown $K$ and (non-constant) exponential decay amplitude functions $A_k(t)$, but again for the stationary setting, were introduced in \cite{MUSIC} and \cite{ESPRIT}, called MUSIC and ESPRIT algorithms, that have profound impact to the current advancement of wireless communication systems, particularly for smart antenna design. Later, the notion of ``synchrosqueezed transform (SST)" was introduced in  \cite{Daub_Maes96} to extend Prony's method to the general non-stationary setting. An extensive study of SST, by using the continuous wavelet transform (CWT) as well as short-time Fourier transform (STFT) constitutes the bulk of the Princeton PhD thesis \cite{Wu_thesis}. Detailed and in-depth early work are developed in \cite{Daub_Lu_Wu11} and \cite{Thakur_Wu11}, on CWT SST and STFT SST, respectively. Later development on SST includes 
\cite{Flandrin_Wu_etal_review13}-\cite{LJL21}. 
 
It is important to point out the objective of signal separation to solve an inverse problem which is quite different from EMD. EMD is an ad hoc computational scheme for decomposing a non-stationary signal into its ``IMFs" and ``trend", without the concern of recovering the true IMF and trend of the source signal. See  \cite{Flandrin04}-\cite{HM_Zhou20}   
for further development of EMD. 
To avoid ``artifacts" aroused by EMD, the empirical wavelet transform based on Fourier-Bessel series expansion was introduced in \cite{Bhattacharyya18}, 
which decomposes multicomponent signals via wavelet based filter banks.  Furthermore, other modified iteration methods, such the variational mode decomposition and the discrete energy separation algorithms in \cite{Upadhyay20} and \cite{Pachori10}, the eigenvalue decomposition algorithms of Hankel matrix in \cite{Jain15} and \cite{Sharma18} are proved to be very efficient for decomposing multicomponent non-stationary signals under some conditions.

To obtain IMFs, the SST method consists of two steps. IFs are estimated from the SST plane.  After recovering of IFs, the IMFs of the source signal are computed by reversible transforms along each estimated IF curves on the SST plane. 

On the other hand, \cite{Chui_M15} introduced a new method with signal separation operator (SSO) 
to solving the inverse problem of multicomponent signal separation. 
 The time-frequency ridge of STFT  (spectrogram) and time-scale ridge of CWT (scalogram) have been studied  
in some papers  including \cite{Stankovic01a, Stankovic08}.  
However those papers are mainly about using ridges as an estimator of IFs.   The ridges of spectrogram or scalogram have not been used directly for multicomponent signal component recovery (mode retrieval) in those papers.  To our best knowledge,  \cite{Chui_M15}  is the first paper to 
 use spectrogram ridges directly for component recovery: a component is recovered simply  by plugging  the ridge to SSO (a variant of STFT). Thus SSO avoids the second step of the two-step SST method in signal separation, which depends heavily on the accuracy of the estimated IFs. 
 In addition,   \cite{Chui_M15} provides theoretical analysis of the error bounds for IF estimation and component recovery.

A window function  is used in SSO.  The window function in \cite{Chui_M15} has the same window length for all the components (sub-signals).  
In this paper we introduce an adaptive signal separation operator (ASSO) which has a time-varying window length. In addition,  the SSO component recovery formula in  \cite{Chui_M15} was derived based on 
the sinusoidal signal approximation, which requires the IFs of component change slowly. Another  
objective of this paper is to deal with signals with fast-varying IFs. To this regard, we consider 
a linear chirp-based model with components of a multicomponent non-stationary signal approximated by linear chirps. 

In this paper we will derive a more accurate component recovery formula based on linear chirp local approximation at any time instant. Furthermore, we propose a signal separation scheme by  adopting a time-varying window for adaptive separation of each sub-signal based on the introduced ASSO and more accurate component recovery formula.  
The main innovations of this paper are: 
(a) we proposed a  more accurate component recovery formula based on linear chirp local approximation; 
(b) we proposed a ridge detection method and a recovery scheme to extract the signal components one by one, and the time-varying window is updated for each component; and (c) the proposed separation algorithm is capable of separating much complicated multicomponent signals and reconstructing the signal trend directly. 
In addition, the proposed method is suitable for engineering implementation with truncated Gaussian window and fast Fourier transform (FFT).   

The remainder of the paper is organized as follows. 
 In Section II, we first review the SSO. After that we formulate and state the results on the ASSO. Finally we show the relationship between the SSO and the STFT with a time-varying parameter $\gs(t)$. In Section III, 
based on the adaptive STFT and  linear chirp local approximation, 
we derive a more accurate component recovery formula
by analyzing the component recovery error 
when the window function is the Gaussian function.  
We propose a signal separation scheme which extracts the signal components one by one with the time-varying window updated for each component. 
  In Section IV,  we present the numerical experiments on synthetic data and real data.
Our experimental results show that ASSO outperforms the EMD, SSO and SSTs in signal separation. 
Finally, we give a brief conclusion in Section V.

\section {Adaptive signal separation operator (ASSO)}
In this section, first we recall the signal separation condition based on the sinusoidal signal local  approximation and review the SSO method. After that we introduce the adaptive signal separation operator with a time-varying parameter. Finally we show the relationship between the ASSO and the adaptive STFT considered in \cite{LCHJJ18}.

\subsection{Signal separation by time-frequency analysis}

The modified STFT of a signal $x(t) \in L_2(\RR)$ with a window function $h(t) \in L_2(\RR)$ is defined by
\begin{equation}
\label{def_STFT}
V_x(t, \eta):=\int_{\RR} x(\tau) h(\tau-t) e^{-j2\pi \eta(\tau-t)}d\tau,
\end{equation}
where $t$ and $\eta$ are the time variable and the frequency variable respectively.
For a real-valued  window function $h(t)$ with $h(0)\ne 0$,  one can show that a real-valued signal $x(t)$ can also be recovered back from its STFT $V_x(t, \eta)$ with integrals involving only $\eta$: 
\begin{equation}
\label{rec_real_x_def}
x(t)=\frac 2{h(0)} {\rm Re}\Big\{ \int_{-\infty}^\infty V_x(t, \eta) d\eta\Big\}. 
\end{equation}

When all components  $x_k(t)$ are sinusoidal signals: $x_k(t) = A_k \cos (2\pi c_k t)$  for some constant  $A_k, c_k>0$,  then for $\eta>0$,  the STFT $V_{x_k}(t, \eta)$ of $x_k(t)$ with a window function $h(t)$ is  
\begin{eqnarray*}
V_{x_k}(t, \eta) \hskip -0.6cm &&= \frac 12 A_k \big(e^{j2\pi c_k t}  \wh h(\eta-c_k)+e^{-j2\pi c_k t}  \wh h(\eta+c_k)\big)\\
&&\approx \frac 12 A_k e^{j2\pi c_k t}  \wh h(\eta-c_k), 
\end{eqnarray*}
provided that  $\wh h$,  the Fourier transform of $h(t)$, decays fast as $\eta\to \infty$.
Moreover, when  $x_k(t)=A_k(t)\cos(2\pi \phi_k(t)) $ with $A_k(t)>0$, $\phi'_k(t)>0$ in \eqref{AM-FM} is well approximated by sinusoidal functions in a small neighborhood of a fixed $t \in \RR$, that is,  
\begin{equation}
\label{sinu_model}
x_k(t+u) \approx A_k(t) \cos \big( 2\pi (\phi_k(t) + \phi'_k(t) u) \big)
\end{equation}
for small $u$, then we have, for $\eta > 0$,
$$
V_{x_k}(t, \eta) \approx \frac 12 A_k(t) e^{j2 \phi_k(t)}  \wh h(\eta-\phi_k'(t)).
$$

Since the support zone of $V_{x_k}(t, \eta)$ for $\eta>0$  is determined by the support of $\wh h$, we need to define the {\it essential support} of $\wh h$ if $h$ is not band-limited. More precisely, for a given threshold $0<{\tau_0}<1$ small enough, if
$|\wh h(\xi)|/\max_\xi|\wh h(\xi)|\le {\tau_0}$ for $|\xi|\ge \lambda_h$, then we say $\wh h(\xi)$ is {\it essentially supported} in $[-\lambda_h, \lambda_h]$, which is denoted as supp($\wh h)\dot{\subseteq}  [-\lambda_h, \lambda_h]$. Note that $\lambda_h=\lambda_{h, {\tau_0}}$ depends on ${\tau_0}$. For simplicity, here and below we drop the subscript ${\tau_0}$.

Hence if  supp($\wh h)\dot{\subseteq}  [-\lambda_h, \lambda_h]$ for some $\lambda_h>0$, then 
$ V_{x_k}(t, \eta)$ essentially lies in the time-frequency zone given by 
$$
Z_k:=\{(t, \eta): |\eta-\phi_k'(t)|< {\lambda_h}, t\in \RR\}.   
$$
Therefore, if 
\begin{equation}
\label{freq_resolution}
\phi'_k(t)-\lambda_h > \phi'_{k-1}(t)+\lambda_h,  2\le k\le K,
\end{equation}
then $Z_k\cap Z_{\ell}=\O, k\not=\ell$, which means the components of $x(t)$ are well separated in the time-frequency plane. \eqref{freq_resolution} is a required condition for the study of FSST in \cite{MOM14} and for the study of the second-order FSST in  \cite{MOM15}. 
This was also pointed in \cite{IMS15}, namely if the STFTs of two components are mixed, the corresponding FSSTs will not be able to separate these two components too. This is also true for other linear time-frequency analysis methods, such as continuous wavelet transform (CWT) and CWT-based SSTs (see \cite{Daub_Lu_Wu11,IMS15}).

\cite{Chui_M15} introduced signal separation operator (SSO) for signal  IF estimation and component recovery of multicomponent signals. More precisely, let $h(t)$ be an admissible window function which is non-negative and even on $\RR$, in $C^3(\RR)$, supp$(h)\subseteq [-1, 1]$ and $h(t)\not \equiv 0$.  
For $a>0$, denote  
\begin{equation}
\label{h_area}
{\wt h}_a := \sum \limits_{n \in \ZZ} h \big(\frac n a\big) .
\end{equation}
When $a$ is large enough, then ${\wt h}_a>0$. The SSO $T^{a,\delta}_x$ of a signal $x(t)$ is defined as
\begin{equation}
\label{def_SSO}
T^{a,\delta}_x(t,\theta) := \frac 1 {{\wt h}_a} \sum \limits_{n \in \ZZ} x(t-n\delta) h \left(\frac n a\right) e^{j2\pi n\theta},
\end{equation}
where $h$ is an admissible window function, $\delta,a>0$ are parameters. 

\cite{Chui_M15} established that, under certain conditions on $A_k(t), \phi_k(t)$,   
when the components of $x(t)$ are well-separated with $T^{a,\delta}_x(t,\theta)$, then the ridge (curve of local maxima) $\widehat \theta_{k}(t)$ of $|T^{a,\delta}_x(t,\theta)|$  corresponding to the IF of the sub-signal $x_k(t)$ gives an approximation to the IF $\phi\rq{}_k(t)$. Most importantly, 
the sub-signal $x_k(t)$ can be recovered directly by
$$\widehat x_k(t) = 2 {\rm Re} \big\{T^{a,\delta}_x(t, \widehat \theta_{k}(t)\big\}.
$$
The reader is referred to \cite{Chui_M15} for the details. In the next two subsections, we introduce adaptive SSO, and then show its relationship to the STFT with a time-varying parameter, termed as the adaptive STFT, considered in \cite{LCHJJ18}.

\subsection {Adaptive signal separation operator (ASSO)}

\label{sec:ASSO}
\begin{mdef}
	\label{ASSO_h} (ASSO).
	Let $x(t)$ be a signal given by \eqref{EMD_model}. 
	For each sub-signal $x_k$ in \eqref{EMD_model}, let $a_{t}>0$ be 
	a time-varying parameter. 
	The (modified) adaptive signal separation operator (ASSO) $\wt T^{a_t,\delta}$  applied to $x$ is defined by 
	\begin{equation}
	\label{def_ASSO_h}
	\wt T^{a_t,\delta}_x(t, \eta) :=  \frac 1 {{\wt h}_{a_{t}}} \sum \limits_{n \in \ZZ} x(t-n\delta) h \big(\frac n {a_{t}}\big) e^{j2\pi \gd n \eta},
	\end{equation}
	where  $h$ is an admissible window function, $ \delta > 0$ and $a_{t}>0$ are parameters, with $a_{t}$ large enough such that ${\wt h}_{a_{t}}$ defined by \eqref{h_area} with $a=a_{t}$ is positive. 
	\end {mdef}  

Note that compared with SSO, ASSO uses  
	$a_{t}>0$ which depends on $t$. 
	In addition, $\gth$ in \eqref {def_SSO} is replaced by $\gd \eta$. Thus the restriction in \cite{Chui_M15} for $\gth$:  $\gth\in [0, 1]$ is replaced by $\eta \in [0, 1/\delta]$.

	In the following we also assume $\int_\RR h(t) dt=1$. In addition, we assume the signals $x(t)$ given by \eqref{EMD_model} satisfy the following conditions, termed as Assumption 1. 
	
	\begin{assum}
		\label{assum:ASSO_condition} 
		For the non-stationary real signal $x=x(t)$ in \eqref{EMD_model}, we assume that $A_0 \in C^0$, $0<A_k \in C^0$, $\phi_k \in C^2$,   and 
		$\phi_k'(t)$ satisfies \eqref{freq_resolution} for $k=1,...,K$, 
		and that there exists  a constant $\vep>0$ such that for small $u$, 
		\begin{equation}
		\label{der_cond}	
		\begin{array}{l}
		|\phi'_k(t+u)-\phi'_k(t)| \le \vep |u | \; \phi'_k(t), \\ |A_k(t+u)-A_k(t)| \le \vep |u| \; A_k(t), \; t\in \RR .
		\end{array}
		\end{equation}
	\end{assum}
	
	Denote 
	$$
	B=B(t):=\max_{1\le k\le K} \phi_k\rq{}(t), \; \mu=\mu(t):=\min_{1\le k\le K} A_k(t). 
	$$
	Then the following theorem for representation and recovery of the $k$-th sub-signal $x_k(t)$ in \eqref{EMD_model} can be followed by Theorem 2.4 in \cite{Chui_M15}.
	
	\begin{theo}
		\label{theo:ASSO_recover_xk} 
		Let $x(t)$ be a  non-stationary signal in \eqref{EMD_model} satisfying Assumption \ref{assum:ASSO_condition}. Let $\gd=\frac 1{\vep a_{t} \sqrt{4 B}}$.  
		Then the following statements hold for sufficiently small $\vep>0$ and fixed $t \in \RR$.

		{\rm (a)} The set $\cG_t:=\big\{ \eta: |\wt T^{a_t,\delta}_x(t,\eta)|>\mu /2\big\}$ is a disjoint union of some non-empty sets $\cG_k:=\cG_{t, k}:=\cG_t \cap \{\eta: (t, \eta)\in Z_k\}$ with $\phi'_k(t)\in \cG_k$. 
		
		{\rm(b)} Let 
		\begin{equation}
		\label{max_theta}
		\wt \eta_k(t):= \arg\max_{\eta \in\mathcal{G}_k}|\wt T^{a_t,\delta}_x(t,\eta)|.  
		\end{equation} 
		Then
		\begin{equation}
		\label{error_theta}
		|\wt \eta_k(t) - \phi'_k(t)| < C \; \vep^{\frac 13}. 
		\end{equation} 
		
		{\rm(c)} With  $\wt \eta_k$ given by \eqref{max_theta}, we have
		\begin{equation}
		\label{error_function}
		\big| 2 {\rm Re} \big\{ \wt T^{a_t,\delta}_x(t, \wt\eta_k(t))\big\} -x_k(t) \big| \le D \; \vep^{\frac 13},   
		\end{equation} 
		where $C$ and $D$ depend on the signal $x(t)$ and the window function $h(t)$. 
	\end{theo}

	\subsection{ASSO and adaptive STFT} 
	\label{sec:relation}
	
	The authors of \cite{LCHJJ18} introduced the adaptive STFT. In the following we show that ASSO is a discretization version of the adaptive STFT. More precisely, let $g=g(t), t\in \RR$ be a 
	window function with $g(0)\not=0$ and having certain smoothness and decaying order as $t\rightarrow \infty$. Denote
	\begin{equation}
	\label{window_general}
	g_\gs(t):=\frac 1\gs g(\frac t\gs),
	\end{equation}
	where $\gs>0$ is a parameter.  For a signal $x(t)$, the STFT of $x(t)$ with a time-varying parameter (termed as the adaptive STFT) is defined in \cite{LCHJJ18} as
	\begin{equation}
	\label{def_STFT_para1}
	\begin{array}{l}
	V^\gs_x(t, \eta) 
	=V^{\gs(t)}_x (t, \eta)
	\\ 
	:=\int_\RR x(\tau)g_{\gs(t)}(\tau-t)e^{-j2\pi \eta (\tau-t)}d\tau
	\\
=\int_\RR  x(t+\tau)\frac 1{\gs(t)}g (\frac \tau{\gs(t)})e^{-j2\pi \eta\tau}d\tau, 
	\end{array}
	\end{equation}
	where $\gs=\gs(t)>0$ is a function of $t$.  
	
	Let us return back to the definition of ASSO. Observe that when $a_{t}$ is large, we have 
	$$
	{\wt h}_{a_{t}}=\sum_{n\in \ZZ} h(\frac n{a_{t}}) \approx a_{t} \int_\RR h(x)d x = a_{t},  
	$$
	where we have used the assumption $\int_\RR h(t)d t = 1$. Thus, 
	\begin{equation*}
	\begin{array}{l}
\wt T^{a_t,\delta}_x(t, \eta) =   \frac 1 {a_{t}} 
	\sum \limits_{n \in \ZZ} x\Big(t-(\delta a_{t})  \frac n{a_{t}}\Big) h\Big(\frac n {a_{t}}\Big) e^{(j2\pi \gd\eta  a_{t}) \frac n {a_{t}}}\\
	 \approx \int_\RR x(t-\delta a_{t} u) h(u) e^{j2\pi \gd\eta  a_{t} u} du\\
	=\int_\RR x(t+\tau) \frac 1{\gd  a_{t}} h\Big(\frac \tau {\gd a_{t}}\Big) e^{-j2\pi \eta \tau } d\tau,  
	\end{array}
	\end{equation*}
	where the last equality follows from the substitution $\tau=-\gd  a_{t} u$ and the fact that $h$ is even. 
	Therefore ASSO $\wt T^{a_t,\delta}_x (t, \eta)$ is a discretization of the adaptive STFT  
	$V^{\gs_t}_x(t, \eta)$ with $g=h$ and $\gs_t=\gd  a_{t}$. 
	
	In the following, when we discuss the IF estimation \eqref{error_theta} and component recovery \eqref{error_function}, we will deal with $V^{\gs_t}_x(t, \eta)$ instead of $\wt T^{a_t,\delta}_x(t, \eta)$. In the next section, we will consider the well-separated condition and more accurate component recovery formula based on the linear chirp local approximation when $g$ is the Gaussian window function.   
	
\section{Component recovery based on ASSO and linear chirp local approximation}
	
As aforementioned, the results with SSO and ASSO in Section II are based on the sinusoidal signal local approximation as given in \eqref{sinu_model}. In this section we study component recovery based on the linear chirp local approximation. Since $\wt T^{a_t,\delta}_x(t, \eta)$ is a variant of the adaptive STFT $V^{\gs_t}_x(t, \eta)$ as shown in Section \ref{sec:relation}, we will focus on $V^{\gs_t}_x(t, \eta)$ from now on due to its simplicity and we also call it ASSO. From now on, unless stated otherwise, we will use $V^{\gs}_x(t, \eta)$ to denote $V^{\gs_t}_x(t, \eta)$ and  we use  the Gaussian function as the window function $g$: 
	\begin{equation}
	\label{def_g}
	g(t)=e^{-\frac {t^2}2}/{\sqrt {2\pi}} . 
	\end{equation}
	
	In this section,  first we consider the well-separated condition. After that we derive a more accurate component recovery formula after analyzing the recovery error. We propose a ridge detection method and an ASSO scheme to recover the component one by one. We also discuss how to select the time-varying parameter $\gs(t)$.
	
	\subsection{Separation condition analysis}
	\label{sec:separt_condi}
	To model a frequency-varying signal more accurately, we consider the local approximation of linear chirps (also called linear frequency modulation signals).
	We say $s(t)$ is a (real) linear chirp if  $s(t)=A \cos \big(2\pi (c t +r t^2/2)\big)$. 
	In this paper we always assume $c+rt>0$ when we talk about a real linear chirp.   Then for $\eta>0$, we have
	\begin{equation}
	\label{STFT_LinearChip}
	V^\gs_s(t, \eta)\approx \frac {A e^{j2\pi(ct +rt^2/2)}}{2\sqrt{1-j2\pi \sigma^2 r}}\;
	\; m\big(\eta-(c+rt)\big), 
	\end{equation}
	where
	\begin{equation}
	\label{m_subfunction}
	m(\xi)= e^{-\frac{2\pi^2 \sigma^2}{1-j2\pi r \sigma^2} \xi^2},  
	\end{equation}
	and throughout this paper, the root $\sqrt{a+ b j }$ of an complex number $a+b j$ with $a>0$ 
	denotes the value locating in the same quadrant as $a+b j$. 
	
To derive \eqref{STFT_LinearChip}, we apply the following formula (see   \cite{table_book} p.121) with $\ga=\sqrt{(2\gs^2)^{-1}-j\pi r}$ and $\gb=2\pi (\eta-c-rt)$:
\begin{equation}
\label{formula_from_table_book}
\int_{-\infty}^\infty e^{-\ga ^2 u^2-i \gb u}du=\frac {\sqrt \pi}\ga e^{-\frac{\gb ^2}{4\ga^2}} 
\; \hbox{(Re($\ga)>0, \gb\in \RR$)}
\end{equation}
to obtain the STFT 	of $\wt s(t)=A e^{j 2\pi (c t +r t^2/2)}$, the complex-version of $s(t)$, is 
	$$
	V^\gs_{\wt s} (t, \eta)=\frac {\wt s (t)}{\sqrt{1-j2\pi \sigma^2 r}} \; m\big(\eta-(c+rt)\big). 
	$$
	Therefore, the STFT 	of $\overline{\wt s}(t)=A e^{-j 2\pi (c t +r t^2/2)}$ is 
	$$
	V^\gs_{\overline{\wt s}}(t, \eta)=\frac {\overline{\wt s} (t)}{\sqrt{1+j2\pi \sigma^2 r}} \; m\big(\eta+(c+rt)\big).  
	$$
	Hence we have  
	$$
	V^\gs_s (t, \eta)=\frac 12 V^\gs_{\wt s} (t, \eta)+\frac 12 V^\gs_{\overline{\wt s}} (t, \eta). 
	$$
	Observe that $|m(\xi)|$ is a Gaussian function and it approaches to 0 very fast at $\xi\to \infty$. In addition, $\eta>0$ and $c+rt>0$, thus the second term on the right-hand side of the above equation is small and could be negligible, and hence, \eqref{STFT_LinearChip} holds.    
	
	Observe that $|m(\xi)|$ gains maximum at $\xi=0$.  
	Thus the ridge of $|V^\gs_s (t, \eta)|$ concentrates around $\eta=c+rt$ in the time-frequency plane.
	Hence if 
	$|m(\xi)|$ is {\it essentially supported} in 
	$[-\lambda_m, \lambda_m]$ for some $\lambda_m>0$, then 
	$V^\gs_s (t, \eta)$  lies in the time-frequency zone given by 
	$$
	\{(t, \eta): |\eta-(c+rt)|< {\lambda_m}, \; t\in \RR\}. 
	$$ 
	
	Recall that we say $|m(\xi)|$ is {\it essentially supported} in $[-\lambda_m, \lambda_m]$ if 
	\begin{equation*}
	|m(\xi)|/\max_\xi|m(\xi)|\le {\tau_0},
	\end{equation*}
	for all $|\xi|\ge \lambda_m$, where  ${\tau_0}$ is a given threshold with $0<{\tau_0}<1$. For $m$ given by \eqref{m_subfunction}, we have 
	\begin{equation}
	\label{m_bandwidth}
	\lambda_m = \sqrt{2|\ln {\tau_0}|} \sqrt{1/{(2\pi \gs)^2} + ( r \gs)^2}. 
	\end{equation}
	
	When  $x_k(t)=A_k(t)\cos(2\pi \phi_k(t)) $ with $A_k(t)>0$, $\phi'_k(t)>0$ in \eqref{AM-FM} is well approximated by linear chirps during any local time of $t \in \RR$, that is,  
	\begin{equation}
	\label{approximation_lfm}
	x_k(t+u) \approx A_k(t) \cos \big( 2\pi (\phi_k(t) + \phi'_k(t) u +\frac {\phi''_k(t) u^2}2) \big)
	\end{equation}
	for small $u$, then 
	following \eqref{STFT_LinearChip}, we have, 
	for $\eta>0$,  
	\begin{equation}
	\label{STFT_LinearChip_general}
	V^\gs_{x_k}(t, \eta)\approx \frac {A_k(t) e^{j2\pi \phi_k(t)}}{2\sqrt{1-j2\pi \sigma^2 \phi''_k(t)}}\;
	e^{-\frac{2\pi^2 \sigma^2(\eta-\phi'_k(t))^2}{1-j 2\pi \phi''_k(t) \sigma^2}}.
	\end{equation}
	Thus $V^\gs_{x_k}(t, \eta)$  lies in the time-frequency zone given by 
	$$
	{\cal Z}_k:= \{(t, \eta): |\eta-\phi'_k(t)|< {\lambda_{k, t}}, \; t\in \RR\},
	$$ 
	where 
	\begin{equation}
	\label{bandwidth_lfm}
	\lambda_{k, t} 
	:= \sqrt{2|\ln {\tau_0}|} \sqrt{1/{(2\pi \gs)^2} + \big( \phi''_k(t) \gs\big)^2}.
	\end{equation}
	Observe that at any (fixed) time instant $t$,  the function on the right-hand side of  \eqref{approximation_lfm} is a linear chirp of variable $u$. When $A_k(t)$ satisfies \eqref{der_cond} for a small $\vep>0$ 
	and $\phi_k^{(3)}(t)$ is small, then \eqref{approximation_lfm} holds (see \cite{CJLS21}). In this case, we say $x_k(t)$ is well-approximated (locally) by linear chirps. 
	
	If $\phi_k(t)$ satisfies 
	\begin{equation}
	\label{freq_lfm_resolution}
	\phi'_k(t)-\phi'_{k-1}(t) > \lambda_{k, t} +\lambda_{k-1,t}, \; t\in \RR,
	\end{equation}
	for all $2\le k\le K$, then the components of $x(t)$ are well separated in the time-frequency plane of STFT, namely, ${\cal Z}_k, 1\le k\le K$, do not overlap. 
	In this case, $x_k(t)$ can be recovered by STFT or the FSST-based methods.
	
	When using the SSO method to estimate IFs and extract sub-signal $x_k(t)$, we care more about the ridges on the time-frequency plane. So the well-separated condition \eqref{freq_lfm_resolution} can be relaxed to 
		\begin{equation}
		\label{freq_lfm_resolution_sso}
		\phi'_k(t) - \phi'_{k-1}(t) > w_{k,t},  \; t\in \RR, 2\le k\le K,
		\end{equation}
		where $w_{k,t}$ is a quantity satisfying   
		\begin{equation}
		\label{separation_degree}
		\max \{\lambda_{k, t}, \lambda_{k-1,t}\} < w_{k,t} < \lambda_{k, t} +\lambda_{k-1,t}. 
		\end{equation}
		For example one may use $ w_{k, t}= \max \{\lambda_{k, t}, \lambda_{k-1,t}\} +\frac 12 \min \{\lambda_{k, t}, \lambda_{k-1,t}\}$. 
		On the one hand $w_{k,t} < \lambda_{k, t} +\lambda_{k-1,t}$ in \eqref{separation_degree}  
		means we allow certain mixture of $x_{k-1}(t)$ and $x_k(t)$ on the STFT plane.
		On the other hand, $w_{k,t} > \max \{\lambda_{k, t}, \lambda_{k-1,t}\}$ means $w_{k,t}$ should be great enough to make sure the values on the ridge (extrema) of $|V^\gs_{x_k}(t, \eta)|$ should not be disturbed by other components and it will not result in new ridges (artifact components) on the STFT plane.

	Note that if $\gs=\gs(t)$ is very small, then $\lambda_{k, t}$ in \eqref{bandwidth_lfm} will be  very large, hence neither \eqref{freq_lfm_resolution} nor \eqref{freq_lfm_resolution_sso} holds. Thus $\gs$ must be larger than a certain positive number, which we denote by $\gs_{\min}$.

	\subsection {More accurate component recovery formula}
	\label{sec:greetings}
	
	As discussed in Section \ref{sec:ASSO}, if the non-stationary signal $x(t)$ in \eqref{EMD_model} satisfies Assumption 1, 	then the components of $x(t)$ can be recovered by SSO and ASSO. In contrast to the (implicit) component recovery error  in \cite{Chui_M15} and  Theorem \ref{theo:ASSO_recover_xk},  we discuss the explicit recovery error with linear chirp local approximation and  the Gaussian  window function given by \eqref{def_g}.    
	We assume the well-separated condition \eqref{freq_lfm_resolution} holds. Thus we can discuss the recovery error for each component individually. For simplicity of presentation, we start with a mono-component signal
	$$
	f(t) = A(t) \cos \left( 2\pi \phi (t) \right). 
	$$  
	Let 
	\begin{equation}
	\label{find_ridge}
	\wc{\eta} (t) := \arg \mathop{\max}\limits_{\eta>0} 
	\left| V^\sigma_f (t,\eta) \right|. 
	\end{equation}
	Then ASSO method tells us that $\wc \eta (t)$ is an estimate of $\phi'(t)$ and  
	$f(t)$ can be recovered by 
	\begin{equation}
	\label{func_recover}
	\wc f(t)  = 2{\rm Re}\big\{V^\sigma_f (t,\wc{\eta}(t))\big\}.
	\end{equation}

	When $f(t)$ can be well-approximated by linear chirps, that is \eqref{approximation_lfm} holds with $x_k(t), \phi'_k(t)$ replaced by $f(t), \phi'(t)$,  then by \eqref{STFT_LinearChip_general}, we have  
	\begin{equation*}
	\begin{array}{l}
	V^{\sigma}_f (t,\wc{\eta}(t)) \approx \frac {A(t) e^{j2\pi \phi(t)}}{2\sqrt{1-j2\pi \sigma^2 \phi''(t)}}\;
	e^{-\frac{2\pi^2 \sigma^2({\small \wc\eta}(t)-\phi'(t))^2}{1-j 2\pi \phi''(t) \sigma^2}}\\
	\approx \frac {A(t) e^{j2\pi \phi(t)}}{2\sqrt{1-j2\pi \sigma^2 \phi''(t)}}\; \hbox{(since $\wc{\eta} (t) \approx \phi\rq{}(t)$)}.
	\end{array}
	\end{equation*}
	Thus the component recovery error is
	\begin{equation}
	\begin{array}{l}
	 \label{func_recover_error}
	{\rm e}_f:= | \wc f(t) -f(t)| 
	\\
	 =\big|2{\rm Re} \big\{ V^\sigma_f (t,\wc{\eta}(t)) \big\}- A(t) \cos \left( 2\pi \phi (t) \right)\big |
	\\ 
	\le \big|2 V^\sigma_f (t,\wc{\eta}(t))- A(t) e^{j  2\pi \phi (t)}\big |
	\\
	\approx  \big| \frac {A(t) e^{j2\pi \phi(t)}}{\sqrt{1-j2\pi \sigma^2 \phi''(t)}} - A(t) e^{j  2\pi \phi (t)}\big |
	\\
= A(t)  \big| \frac 1{\sqrt{1-j2\pi \sigma^2 \phi''(t)}} -1 \big |
	\\ 
	=A(t) 
	\big|  \frac {j2\pi \sigma^2 \phi''(t)}{\sqrt{1-j2\pi \sigma^2 \phi''(t)}(1+\sqrt{1-j2\pi \sigma^2 \phi''(t)})} \big |
	\\ 
	\le   \frac {2\pi \sigma^2 |\phi''(t)| A(t)}{\big(1+4\pi^2 \sigma^4 \phi''^2(t)\big)^{\frac 14}\; \big(1+ \sqrt{1+4\pi^2 \sigma^4 \phi''^2(t)}\big)^{\frac 12}}, 
	\end{array}
	\end{equation} 
	where the last inequality follows from 
	$$
	|\sqrt{1-j2\pi \sigma^2 \phi''(t)}+1| \ge \big(1+ \sqrt{1+4\pi^2 \sigma^4 \phi''^2(t)}\big)^{\frac 12}. 
	$$
	Hence,  ${\rm e}_f$ is essentially bounded by $2\pi \sigma^2 |\phi''(t)| A(t)$. 
	Of course smaller $\gs(t)$ will result in smaller error. Due to that $\gs(t)\ge \gs_{\min}$,  $\wc f(t)$ gives a good approximation to $f(t)$ only if $|\phi''(t)|$ is small, meaning IF of $f$ changes slowly.  It seems we cannot break the bottleneck of slowly changing IFs requirement even if we consider the linear chirp approximation.  
	However, the error analysis in \eqref{func_recover_error} triggers us to propose the following component recovery formula  
	\begin{equation}
	\label{func_recover2}
	\wc f(t) = 2{\rm Re}\Big\{\sqrt{1-j2\pi \gs^2 \phi''(t) } \;  V^\sigma_f (t,\wc{\eta}(t)) \Big\}. 
	\end{equation}
	Then the corresponding recovery error is 
	\begin{equation*} 
	\begin{array} {l}
	|\wc  f(t) -f(t)| 
	\\=\big| 2 {\rm Re} \big\{\sqrt{1-j2\pi \gs^2 \phi''(t) } \;   V^\sigma_f (t,\wc{\eta}(t)) \big\}- f(t)\big |
	\\ 
	\le \big|2 \sqrt{1-j2\pi \gs^2 \phi''(t) } \;  V^\sigma_f (t,\wc{\eta}(t))- A(t) e^{j  2\pi \phi (t)}\big |
	\\ \approx  \big| {A(t) e^{j2\pi \phi(t)}} - A(t) e^{j  2\pi \phi (t)}\big |=0. 	
	\end{array} 
	\end{equation*} 	
	Thus \eqref{func_recover2} provides a more accurate recovery formula. 
	
For a complex-valued signal $f(t)=A(t)e^{j2\pi \phi(t)}$, derived similarly as the above, a more accurate recovery formula is  
	\begin{equation}
	\label{func_recover2_complex}
	\wc f(t) = \sqrt{1-j2\pi \gs^2 \phi''(t) } \;  V^\sigma_f (t,\wc{\eta}(t)). 
	\end{equation}

	Observe that the recovery formula \eqref{func_recover2} 
	involves $\phi''(t)$. In practice, we need to estimate $ \phi''(t)$. One simple way is to use 
	$\wc \eta^\gp(t)$ as $\phi''(t)$ since $\wc \eta(t)$ is an estimate to $\phi\rq{}(t)$.
	One can apply a three-point or five-point formula for differentiation to the ridge $\wc \eta(t)$ to obtain its derivative.  In this paper we use linear fitting of $\wc{\eta}(t)$:  for a fixed $t$, let $\wc r_t$ defined by 
	\begin{equation}
	\label{linear_fitting}
	\min_u\big \| \wc{\eta}(t+u) - \big(\wc{\eta}(t)+\wc r_{t}u\big)\big \|_2
	\end{equation}
where $u$ ranges over $[-\gl_0, \gl_0]$ with $\gl_0$ the essential support of  $g_{\gs}$:
\begin{equation}
\label{def_gl0}
\gl_0:=2\pi \gs \sqrt{2|\ln {\tau_0}|}.
\end{equation}
Then we use $\wc r_t$ as $\phi''(t)$  in \eqref{func_recover2} and \eqref{func_recover2_complex}. 
	
	\bigskip		
	Next let us consider the component recovery of a multicomponent signal $x(t)$ of the form 
	\eqref{EMD_model}. We assume \eqref{freq_lfm_resolution} holds, that is  the components of $x(t)$ are well separated in the time-frequency plane. Denote 
	$$
	\begin{array}{l}
	\cH_t:=\big\{ \eta: |V^\gs_x(t,\eta)|>\mu /2\big\}, \\
	\cH_k=\cH_{t, k}:=\cH_t \cap \{\eta: (t, \eta)\in {\cal Z}_k\}.
	\end{array}
	$$
	When $x_k(t)$ is well-approximated locally by linear chirps, that is \eqref{approximation_lfm} holds, then $\phi'_k(t)\in \cG_k$. Let  
	\begin{equation}
	\label{find_ridge_k}
	\wc{\eta}_k(t) := \arg \mathop{\max}\limits_{\eta\in \cH_k} 
	\left| V^{\sigma}_x (t,\eta) \right|. 
	\end{equation}
	Then $\wc \eta_k(t)$ is an estimate of $\phi'_k(t)$. By the above derivation with the mono-component signal $f$, we propose the following formula  to recover components. 
	
{\bf Component recovery formula based on linear chirp local approximation:} Component  $x_k(t)$ in  \eqref{EMD_model} can be recovered by 
	\begin{equation}
	\label{func_recover2_k}
	\wc x_k(t) = 2{\rm Re} \Big\{\sqrt{1-j2\pi \gs^2 \phi''_k(t) } \;  V^{\sigma}_x (t,\wc{\eta}_k(t)) \Big\}; 
	\end{equation}
if $x_k(t)$ in \eqref{EMD_model} is a complex signal, namely $x_k(t)=A_k(t)e^{j2\pi\phi_k(t)}$,  
then the recovery formula is  
	\begin{equation}
	\label{func_recover2_complex_k}
	 \wc x_k(t) = \sqrt{1-j2\pi \gs^2 \phi''_k(t) } \;  V^{\sigma}_x(t,\wc{\eta}_k(t)). 
	\end{equation}

Again, to apply \eqref{func_recover2_k}  or \eqref{func_recover2_complex_k},  we need to have an estimation of  $\phi''_k(t)$. As above, we may use   $\wc \eta'_k(t)$ as $\phi''_k(t)$, or 
	use the linear fitting 
	\begin{equation}
	\label{linear_fitting_k}
	\min_u\big\| \wc{\eta}_k(t+u) - \big(\wc{\eta}_k(t)+\wc r_{t, k}u\big) \big\|_2
	\end{equation}
	to obtain $\wc r_{t, k}$ as an approximation of  $\phi''_k(t)$. 	
	
	Next we consider the recovery of the (real-valued) trend $A_0(t)$. Assume 	$\phi'_1(t) $ satisfies 
	$$\phi'_1(t) >\gl_{1,t}+ \gl_{0}, 
	$$   
	where $\gl_{0}$ is defined by \eqref{def_gl0}.  Then $A_0(t)$ and $x_1(t)$ are well-separated in the adaptive STFT plane. In this case we use the following formula to recover $A_0(t)$: 
	\begin{equation}
	\label{trend_recover}
	\wc A_0 (t) ={\rm Re} \big\{ V^{\sigma}_x (t,0) \big\}.
	\end{equation}
With the fact 
	\begin{equation*}
	\begin{array}{l}
	\wc A_0 (t) = {\rm Re} \big(\{V^{\sigma}_x (t,0) \big\} \approx  V^{\gs}_{A_0} (t, 0)  
	\\= \int_\RR A_0 (\tau)g_{\gs}(\tau-t)d\tau, 
	\end{array}
	\end{equation*}
we know the recovery error is
	\begin{equation*}
	\begin{array}{l}
	|\wc A_0 (t)-A_0(t)| \approx |\int_\RR A_0 (\tau) g_{\gs}(\tau-t)d\tau - A_0(t) | 
	\\ =  | \int_\RR A_0 (t+\tau) g_{\gs}(\tau)d\tau -A_0(t) |
	\\ =  | \int_\RR ( A_0 (t+\tau) -A_0 (t) )  g_{\gs}(\tau)d\tau |
	\\	 \le \int_\RR  |  A_0 (t+\tau) -A_0 (t)  | g_{\gs}(\tau)d\tau  \\
	\le \int_\RR \epsilon |\tau A_0(t)| g_{\gs}(\tau)d\tau  \\
	= \sqrt{\frac 2 {\pi}} |A_0(t)| \epsilon  \gs.
	\end{array}
	\end{equation*}
	Note that here we assume $\big | A_0 (t+\tau) -A_0 (t)  \big| \le \epsilon |\tau A_0(t)| $, which is consistent with \eqref{der_cond} when $k=0$. Thus the recovery error is small if  $\epsilon$ is small. In addition, a smaller  $\gs$  results in a more accurate recovery.

{\bf Remark 1.} 	
	{\it In the paper we propose component recovery formulas \eqref{func_recover2_k} and  \eqref{func_recover2_complex_k} based on linear chirp local approximation. Compared with 
	sinusoidal signal-based recovery formula \eqref{func_recover}, there is a factor $\sqrt{1-j2\pi \gs^2 \phi''_k(t) }$ in \eqref{func_recover2_k} and \eqref{func_recover2_complex_k}. 
	The experiments in the next section show that \eqref{func_recover2_k} and \eqref{func_recover2_complex_k} lead to more accurate recovery 
	results than those by \eqref{func_recover}  even when the estimate of $\phi''_k(t)$ is rough.  
	More mathematically rigorous  analysis for the recovery error by \eqref{func_recover2_complex_k} 
	was studied very recently by authors of this paper and their collaborator in \cite{CJLL20}. } 
	
	In Section \ref{sec:iASSO} we will propose an iterative ASSO scheme based on \eqref{func_recover2_k} in which component $x_k(t)$ is recovered one by one. In this case, the time-varying parameter $\gs(t)$ is chosen for this particular $x_k(t)$, and hence, it is denote by 
	$\gs_{k, t}$. The choice of $\gs_{k, t}$ will be discussed in Section \ref{sec:choose_gs}. 
	
Before moving on to the next subsection, we remark that during the review process of this paper,  we were aware of the paper \cite{LM1} which was submitted to a journal in April 2020, more than two months later than our paper submitted to this journal. 
	Let $f$ be the complex version of \eqref{AM-FM}, namely $f(t)=\sum_{k=1}^K f_k(t)$ with 
	$f_k(t)=A_k(t)e^{j2\pi \phi_k(t)}$. The  authors of \cite{LM1} proposed to approximate the STFT $V^{\wt g}_{f_k}$ of $f_k(t)$  by the following formula
	\begin{equation}
	\label{STFT_their}
	V^{\wt g}_{f_k}(t, \eta)\approx V^{\wt g}_f\big(t, \wh \go_f^{[2]}(t, \varphi_k(t))\big)e^{-\frac{\pi\gs^2 \big(\eta-\wh \go_f^{[2]}(t, \varphi_k(t))\big)^2}{1-j\wh q_f(t, \varphi_k(t)\gs^2}}
	\end{equation}
	where $V^{\wt g}_{f}(t, \eta)$ denotes the STFT defined by \eqref{def_STFT} with window function given by 
$$\wt g(t)=e^{-\frac {t^2}{(\gs/\sqrt{\pi})^2}},$$
	$\varphi_k(t)$ is a ridge corresponding to the $k$-th component $x_k$, like $\wc \gth_k(t)$ defined by in \eqref{find_ridge_k}, $\wh \go_f^{[2]}(t, \eta)$ and $\wh q_f(t, \eta)$ are quantities used to define the 2nd-order FSST. The authors claim that $\wh \go_f^{[2]}(t, \varphi_k(t)) \approx \phi'_k(t), 
	q_f(t, \varphi_k(t))\approx \phi''_k(t)$.  Thus the recovered $f_k(t)$ proposed in \cite{LM1} is  
	\begin{eqnarray}
	\nonumber &&f_k(t)=\frac 1{\wt g(0)}\int_{-\infty}^\infty V^{\wt g}_{f_k}(t, \eta)d\eta \\
	\nonumber && \approx \int_{-\infty}^\infty V^{\wt g}_f\big(t, \wh \go_f^{[2]}(t, \varphi_k(t))\big)e^{-\frac{\pi\gs^2 \big(\eta-\wh \go_f^{[2]}(t, \varphi_k(t))\big)^2}{1-j\wh q_f(t, \varphi_k(t)\gs^2}} d\eta\\
	\nonumber && =  V^{\wt g}_f\big(t, \wh \go_f^{[2]}(t, \varphi_k(t))\big)
	\int_{-\infty}^\infty e^{-\frac{\pi\gs^2 \eta^2}{1-j\wh q_f(t, \varphi_k(t)\gs^2}} d\eta\\
	&& \label{recovery_their}
	=  \frac 1{\gs} \sqrt{1-j\gs^2\wh q_f\big(t, \varphi_k(t)\big)}\;  V^{\wt g}_f\big(t, \wh \go_f^{[2]}(t, \varphi_k(t))\big), 
	\end{eqnarray}
where the last equality follows from \eqref{formula_from_table_book} with 
$\ga=\gs \sqrt \pi/ \sqrt{1-j\gs^2\wh q_f\big(t, \varphi_k(t)\big)}$ and $\gb=0$. 
Note that $\wt g(t)=\gs g_{\gs/{\sqrt \pi}}(t)$. Thus 
	\eqref{recovery_their} is exactly \eqref{func_recover2_complex_k}, the complex version of 
	our recovery formula   \eqref{func_recover2_k}, 
	with window function $\wt g$, and  $\wh q_f(t, \varphi_k(t))$  and $\wh \go_f^{[2]}(t, \varphi_k(t))$ used for $\phi''_k(t)$ and $\wc\gth_k(t)$ respectively. 
	The same  authors of \cite{LM1} considered in their very recent paper \cite{LM2} another recovery formula ( see (27) in \cite{LM2}) which is the discrete form of  
	\begin{equation}
	\label{recovery_theri2} 
	f_k(t)\approx  \frac 1{\gs} \sqrt{1-j\gs^2 \big(D_k^{fin}\big)\rq{}(t) }\;  V^{\wt g}_f\big(t, D_k^{fin}(t) )\big), 
	\end{equation}
	where $ D_k^{fin}(t)$ is a ridge. Again \eqref{recovery_theri2} is our recovery formula   \eqref{func_recover2_complex_k} with $\wc\gth_k '(t)$ used as an approximation to $\phi''_k(t)$. 
	
	\subsection {Component recovery formula with iterative ASSO}
	\label{sec:iASSO}
	
	In the above subsection we derive more accurate component recovery formulas \eqref{func_recover2_k} and \eqref{func_recover2_complex_k}. 
	In this subsection, we propose, with these formula, to reconstruct the component of $x(t)$  one by one, to say to recover $x_\ell(t)$ first, then a different component $x_m(t)$ of $x(t)$ from $x(t)-x_\ell(t)$, 
	and so on. When we target a particular component, to say $x_k(t)$, we will choose $\gs(t)$, denoted by $\gs_{k,t}$ depending only on IFs and their derivatives of $x_{k-1}, x_k, x_{k+1}$ such that $x_{k-1}, x_k, x_{k+1}$ are well-separated in the adaptive STFT (ASSO) plane. More precisely, we will choose 
	$\gs_{k}$ such that 
$ V^{\gs_{k,t}}_{x_{k-1}}(t,\eta)$,  $V^{\gs_{k,t}}_{x_k}(t,\eta)$, $V^{\gs_{k,t}}_{x_{k+1}}(t,\eta)$ lie in non-overlapping time-frequency zones, which can be guaranteed by 
		\begin{equation}
		\label{separate_kth_component_sso}
		\phi'_k(t) - \phi'_{k-1}(t) >\gl_{k,t}+ \gl_{k-1,t}, \phi'_{k+1}(t) - \phi'_k(t) > \gl_{k+1,t}+\gl_{k,t} 
		\end{equation}
		where $t\in \RR$, $\gl_{k, t}$ is defined by \eqref{bandwidth_lfm}. 
		Note that for $k=K$, only one inequality in \eqref{separate_kth_component_sso} is required; for $k=1$, $x_1(t)$ and the trend $A_0(t)$ should be well-separated on the ASSO plane if 
		$$\phi'_1(t) >\gl_{1,t}+ \gl_{0}, \; \phi'_2(t) - \phi'_1(t) > \gl_{2,t}+\gl_{1,t}, \; t\in \RR$$  
		where $\gl_{0} = \gl_{g_{\sigma}}$.
		With the analysis above, in particular by \eqref{func_recover2}, 
		we propose the following recovery formula for $x_k(t)$: 
		\begin{equation}
		\label{ASSO_recon_generalized}
		\wc x_k (t) = 2{\rm Re} \big\{ \sqrt{1-j2\pi \gs_{k,t}^2 \wc r_{k,t} } V^{\gs_{k,t}}_x (t,\wc \eta_{t})\big\}, 
		\end{equation}    
		where	\begin{equation}
		\label{theta_r}
		\wc \eta_{t} = \arg\max_{\eta\in\cH_k}| V^{\gs_{k, t}}_x (t,\eta)|, 
		\end{equation}
		and $ \wc r_{k,t} $ is the estimation of chirp rate $\phi''_k(t)$
		obtained by \eqref{linear_fitting} with $\wc \eta (t)$ replaced by $\wc \eta_{t} $ and $u$ ranging over $[-2\pi \gs_{k,t} \sqrt{2|\ln {\tau_0}|}, 2\pi \gs_{k,t} \sqrt{2|\ln {\tau_0}|}]$. 
	
We should choose the parameter $\gs(t)$ as the minimum $\sigma$ for each time $t$, with which $x_k$ is well-separated from $x_{k-1}$ and $x_{k+1}$ 
on the adaptive STFT plane, namely the inequalities  in \eqref{separate_kth_component_sso} hold. 
We propose a method (called Algorithm 2) to choose $\gs_{k, t}$ in the next subsection. 

Next we present a signal separation scheme to recover components one by one. 
First, we consider the trend of the input signal $x(t)$ with a small constant $\gs$, with which the trend is well-represented along $\eta=0$. We extract $A_0(t)$ by \eqref{trend_recover}.  After that we consider the component with the largest peak of the trend-removed signal $s(t)$:
$$
s(t)=x(t)-\wc A_0(t). 
$$
Before we describe our procedure, noting that in practice we do not know or it is hard to estimate $\cH_k$, first we need a method to obtain the ridge or an estimate of it. In this paper we propose a method as follows.   
Suppose  $s(t)$ is uniformly sampled with sampling points $t_n, n=1, 2, \cdots, N$. 
We let $T$ and $\varUpsilon$ denote the sample points:
$$
T=\{t_1,t_2, \cdots, t_N\}, \;\;  \varUpsilon =\{\eta_1,\eta_2, \cdots, \eta_{N_1}\}, 
$$
where $N$ and $N_1$ are the number of time and frequency points respectively.

 Let  $V^{\gs}_s (t,\eta)$ be the adaptive STFT of $s(t)$ with time-varying $\gs(t)$. 
 We find the maximum points 
 \begin{equation}
 \label{def_peak}
(t_m,\eta_m):= \arg \max_{t\in T, \eta>0} \big| V^{\gs}_s (t,\eta)\big|,  
\end{equation}
Next we define 
$$
\eta_{m+1}:= \arg \max_{\eta\in \cD_{t_m}} \big| V^{\gs}_s (t_{m+1},\eta)\big|, 
$$
where 
$$
\cD_{t_m}:=[\eta_m-\gl_0, \eta_m+\gl_0]
$$
with $\gl_{0}$ defined by \eqref{def_gl0}. From $\eta_{m+1}$, we define $\cD_{t_{m+1}}$ and then obtain $\eta_{m+2}$ and so on. We obtain $\eta_{m-1}$, then $\eta_{m-2}$ so on in the same way.  More precisely, we propose the following ridge detection algorithm.

\n {\bf Ridge Detection Algorithm}

\n Step 1. Obtain $t_m,\eta_m$ by \eqref{def_peak}. 

\n Step 2. For $n=m, m+1, \cdots, N-1$, do
$$
\cD_{t_n}=[\eta_n-\gl_0, \eta_n+\gl_0], \; \eta_{n+1}= \arg \max_{\eta\in \cD_{t_n}} \big| V^{\gs}_s (t_{ n+1},\eta)\big|;
 $$
 and for $n=m, m-1, \cdots, 2$, do
$$
\cD_{t_n}=[\eta_n-\gl_0, \eta_n+\gl_0], \; \eta_{n-1}= \arg \max_{\eta\in \cD_{t_n}} \big| V^{\gs}_s (t_{n-1},\eta)\big|. 
 $$

\n Step 3. Obtain ridge $ \wc\eta_1 (t_n)=\eta_n,  n=1, \cdots, N$.  
\hfill $\square$
 
 \medskip 
 
 Let us return back to the trend removed multicomponent signal $s(t)$. After obtaining the ridge $\wc \eta_1 (t), t\in T$ by the ridge dectection algorithm, we  estimate $ \wc r_{1,t}$ by \eqref{linear_fitting} with $\wc \eta(t)= \wc {\eta}_1(t)$ and recover one component, denoted by $f_1(t)$ given by 
 $$
 f_1(t) =2 {\rm Re} \Big \{\sqrt{1-j2\pi \gs^2 \wc r_{1,t} } V^{\gs}_s (t, \wc \eta_1 (t))\Big \}.
 $$
 After that, we consider $s(t)-f_1(t)$, and repeat the above process to obtain $\wc \eta_2 (t)$ and $f_2(t)$; and then $\wc \eta_3 (t)$ and $f_3(t)$ and so on until we obtain all components with a possible residual $d(t)$ of $x(t)$. 
 Thus we have 
 \begin{equation}
	\label{reconstruction_result}
	\wc x (t)= \wc A_0(t)+\sum_p f_p(t) +d(t).
	\end{equation}
In Algorithm 1, we provide the pseud-codes for the detailed iteration process. Note that  Steps 3-13 in Algorithm 1 are the ridge detection algorithm proposed above including  
 $\gamma_1$ and $\gamma_2$ as thresholds for the sake of noise. Here we let $\gamma_2 < \gamma_1$. 
We also smooth $\sigma _{\rm R}(t)$ and $\wc r_{p,t}$ in Algorithm 1 with some low-pass filters. 
 In addition, observe that the duration of a component $T_p, p=1,2,3,... $ resulted from Algorithm 1 may different from another. 	
 
	In Algorithm 1, we call $\gs_{\rm R}(t)$ the global optimal time-varying parameter, with which all components are expect to be well-separated in the ASSO plane. 
	To further increase the recovery accuracy, we define $\gs_{p,t} = \gs_{p}(t)$ as the local optimal time-varying parameter, with which the component  corresponding to $\wc \eta_p( t)$ 	are expect to be well-separated from other components. We will provide the methods to estimate $\gs_{p}(t)$ in the following subsection. 

\begin{algorithm}[ht]
	\caption  {\; Adaptive recovering process}
	\begin{algorithmic}  
		\STATE 1.\; \ {\bf Input}: Uniform samples $s^{(p)}(t)=s(t)$ with $t \in T = \{t_1,t_2,...,t_N\}$, $\triangle t = t_2-t_1$ and $p = 1$.
		\STATE 2.\; \ Calculate  $\gs = \gs_{\rm R}(t)$ for $s^{(p)}(t)$ with \eqref{def_gs_global} in \S\ref{sec:choose_gs}.
		\STATE 3.\; \ Let  $(t_{\rm m},\eta_{\rm m}) = \arg \max \big| V^{\gs}_{s^{(p)}} (t,\eta)\big|$, $ \wc {\eta}_{p}(t_{\rm m}) = \eta_{\rm m}$. 
		\STATE 4.\; \ {\bf If} $\big| V^{\gs}_{s^{(p)}} (t_{\rm m},\eta_{\rm m})\big| > \gamma_1$
		\STATE 5.\; \ \ \ \ \ \  Search the ridge towards the right. Let $t \leftarrow t_{\rm m}+ \triangle t $,  \\ 
		\; \ \ \ \ \ \ \ ${\eta}_{\rm m} = \arg \mathop{\max}_{\eta \in \cD_{t-\triangle t}} \left| V^{\gs}_{s^{(p)}} (t,\eta) \right|$. \;\;\;\; {\textcircled 1}
		\STATE 6. \;\ \ \ \ \    {\bf While} $\big| V^{\gs}_{s^{(p)}} (t,\eta_{\rm m})\big| > \gamma_2$, 
		\STATE 7. \;\ \ \ \ \ \ \ \ \  \ $t \leftarrow t + \triangle t $, 
		\STATE 8. \;\ \ \ \ \ \ \ \  \ \ Update $\eta_m$ by {\textcircled 1 }, and let $\wc {\eta}_{p}(t) = \eta_{\rm m}$. 
		\STATE 9.\; \ \ \ \ \  Search the ridge towards the left. Let $t \leftarrow t_{\rm m} - \triangle t $, \\   
		\; \ \ \ \ \ \ \ ${\eta}_{\rm m} = \arg \mathop{\max}_{\eta \in \cD_{t+\triangle t}} \left| V^{\gs}_{s^{(p)}} (t,\eta) \right|$.\;\;\;\; \; {\textcircled 2}
		\STATE 10. \;\ \ \ \   {\bf While} $\big| V^{\gs}_{s^{(p)}} (t,\theta_{\rm m})\big| > \gamma_2$, 
		\STATE 11. \;\ \ \ \ \ \ \ \ \   $t \leftarrow t - \triangle t $, 
		\STATE 12. \;\ \ \ \ \ \ \ \ \   Update $\eta_m$ by {\textcircled 2 }, and let $\wc {\eta}_{p}(t) = \eta_{\rm m}$. 
		\STATE 13. \;\ \ \ \  Estimate $\wc r_{p,t}$ with \eqref{linear_fitting} with $\wc \eta(t)=  \wc {\eta}_{p}(t)$.
		\STATE 14. \;\ \ \ \ \ $f_p(t) =2 {\rm Re}  \big\{\sqrt{1-j2\pi \gs^2 \wc r_{p,t} } V^{\gs}_{s^{(p)}} (t,\eta_{p,t}^{\rm r})\big \}$.
		\STATE 15. \;\ \ \ \ \  $s^{(p+1)}(t) = s^{(p)}(t)-f_p(t)$, $t\in T_p$.
		\STATE 16. \;\ \ \ \ \  $p \leftarrow p+1$, goto Step 2.
		\STATE 17. \; {\bf Else} $ d(t) = s^{(p)}(t)$.
		\STATE 18. \; {\bf End if}.
		\STATE 19. \; {\bf Outputs:\;} components $\{f_p(t), p=1,2,3,...\}$ and the residual $d(t)$.
	\end{algorithmic}
\end{algorithm}

	\subsection {Estimation of the time-varying parameter}
	\label{sec:choose_gs}
	When we recover a particular component $x_k(t)$, we assume  \eqref{separate_kth_component_sso} holds for some $\gs=\gs_k(t)$. 
If we know $\phi'_\ell(t)$ and $\phi''_\ell(t)$, for $\ell=k-1,k, k+1$, 
then we choose a minimum $\gs_k(t)$ with \eqref{separate_kth_component_sso} holding true so that we have an accurate recovery of $x_k(t)$. However in practice, we in general have no prior knowledge of $\phi'_k(t)$ and $\phi''_k(t)$. Hence, we need to have a method to obtain a suitable $\gs_k(t)$. 
	
	First, we use the R${\rm \acute e}$nyi entropy to obtain a preliminary choice of $\gs(t)$. 
	The R${\rm \acute e}$nyi entropy is a commonly used measurement to evaluate the concentration of a time-frequency representation such as STFT in \eqref{def_STFT}, Wigner-Ville distribution \cite{Leon_Cohen} (WVD), SST etc. of a signal of $x(t)$, see \cite{Baraniuk01, Stankovic01,  Wu17, SP18, LCJ18}.
	
	In this paper, we define the local R${\rm \acute e}$nyi entropy for 
	$V^{\gs}_x (t,\eta) $ as
	\begin{equation}
	\label{def_renyi_entropy_global}
	\begin{array}{l}
	E_{\zeta, \gs} (t) : = 5 \log_2 {\int_{t - \zeta }^{t + \zeta }
		{\int_0^\infty  {\left| V^{\gs}_x (b,\eta)\right|^2 d\eta db} }}
	\\
	\qquad \qquad -2 \log_2 \int_{t - \zeta }^{t + \zeta } {\int_0^\infty  {|V^{\gs}_x (b,\eta)|^5 d\eta db} },
	\end{array}
	\end{equation}
	where $\zeta$ is the localization parameter to determine the length of local time. 
	One may refer to \cite{Stankovic01} for other measures of time-frequency concentrations. 
	Note that the smaller the R${\rm \acute e}$nyi entropy, the better the time-frequency resolution. So for a fixed time $t$ and parameter $\zeta$, we use \eqref{def_renyi_entropy_global} to find a $\gs_{\rm R}$ (denoted by $\gs_{\rm R}(t)$) such that  $V^{\gs_{\rm R}}_x (t,\theta)$ has the best time-frequency concentration, namely,  
	\begin{equation}
	\label{def_gs_global}
	\gs_{\rm R} (t):= \mathop {\rm argmin }\limits_{\sigma > 0}  \left\{ E_{\zeta, \gs} (t) \right\}.
	\end{equation}
	We call	$\gs_{\rm R}(t)$ the global optimal time-varying parameter. 
	
	Next, for a fixed $t$, we consider a smaller $\gs(t)$, to say $\gs_{\rm R}(t)-\gt$, where $\gt>0$ denotes a decrement for $\sigma$. Then we consider 
	the difference between the original and the updated ridges 
	$\big| \eta_p^{(q)}(t) - \eta_p^{(q+1)}(t)\big|$. If the difference is small enough, we consider a further smaller $\gs(t)$ which is $\gs_{\rm R}(t)-2\gt$. 
	$E_r$ is a threshold to measure the difference, which is related to $\lambda_{p, t}$. One may choose $E_r = \epsilon \lambda_{p, t} $, where $\epsilon \in (0,0.1)$.     
	The detailed procedure is  provided in Algorithm 2 (for a fixed $t\in T_p$).

	\begin{algorithm}[ht]
		\caption  {\; Searching for the local optimal parameter}
		\begin{algorithmic}  
			\STATE 1.\;\ {\bf Input}: $s^{(p)}(t)$, $\eta_p(t) = \wc \eta_{p,t} $ and $\gs_{\rm R}(t)$, $t\in T_p$ from Algorithm 1. 
			\STATE 2.\;\ Initialize $q = 1$, $\eta_p^{(q)}(t) = \eta_p(t)$, $\gs^{(q)}(t) = \gs_{\rm R}(t)$.
			\STATE 3.\;\ Let $\gs^{(q)}(t) \leftarrow \gs^{(q)}(t)-\gt_\sigma$. 
			\STATE 4.\;\  $\eta_p^{(q+1)}(t) = \arg \mathop{\max}_{\eta \in \cH_{t}} \big|V^{\gs^{(q)}} _{s^{(p)}} (t,\eta) \big|$. 
			\STATE 5.\;\ {\bf If} $\big| \eta_p^{(q)}(t) - \eta_p^{(q+1)}(t)\big| < E_r$, 
			\STATE 6. \;\ \ \ \ \ $q \leftarrow q+1$, goto Step 3.
			\STATE 7.\; {\bf End if}.
			\STATE 8.\;\ {\bf Output:\;} $\gs_p(t) = \gs^{(q)}(t)$.
		\end{algorithmic}
	\end{algorithm}

	\section {Numerical experiments}
In this section, we provide some numerical examples to further illustrate the effectiveness and robustness of our method in component recovery. 	

\subsection {Synthetic signal}

First  we consider a mono-component LFM signal to show how the linear chirp local approximation-based SSO formula  \eqref{func_recover2_k} improves the recovery accuracy when compared with the conventional sinusoidal signal local approximation-based SSO method. Let $x(t) = \cos( 34 \pi t + 37 \pi t^2)$, where $t \in [0,1]$ with $512$ sampling points.  
Here we simple let $\gs=0.02$  (a constant).   The left panel of Fig.1 shows the recovery errors 
$|x(t) - \wc x(t)  |$ 
with these two methods, where the ground truth $\phi^{\gp\gp}_k(t)$ is used in  \eqref{func_recover2_k}, while the right panel of Fig.1 is the recovery errors, 
where $\wc r_{t, k}$ is used in  \eqref{func_recover2_k} as an estimate of $\phi^{\gp\gp}_k(t)$. 
Because of the bound effect, here we just consider the center part of the signal, namely $t\in [0.2, 0.8]$.
Observe that the recovery error is dramatically reduced by using the proposed linear chirp model.

\begin{figure}[th]
	\begin{minipage}{0.48\linewidth}
		\centerline{\includegraphics[width=6.6cm]{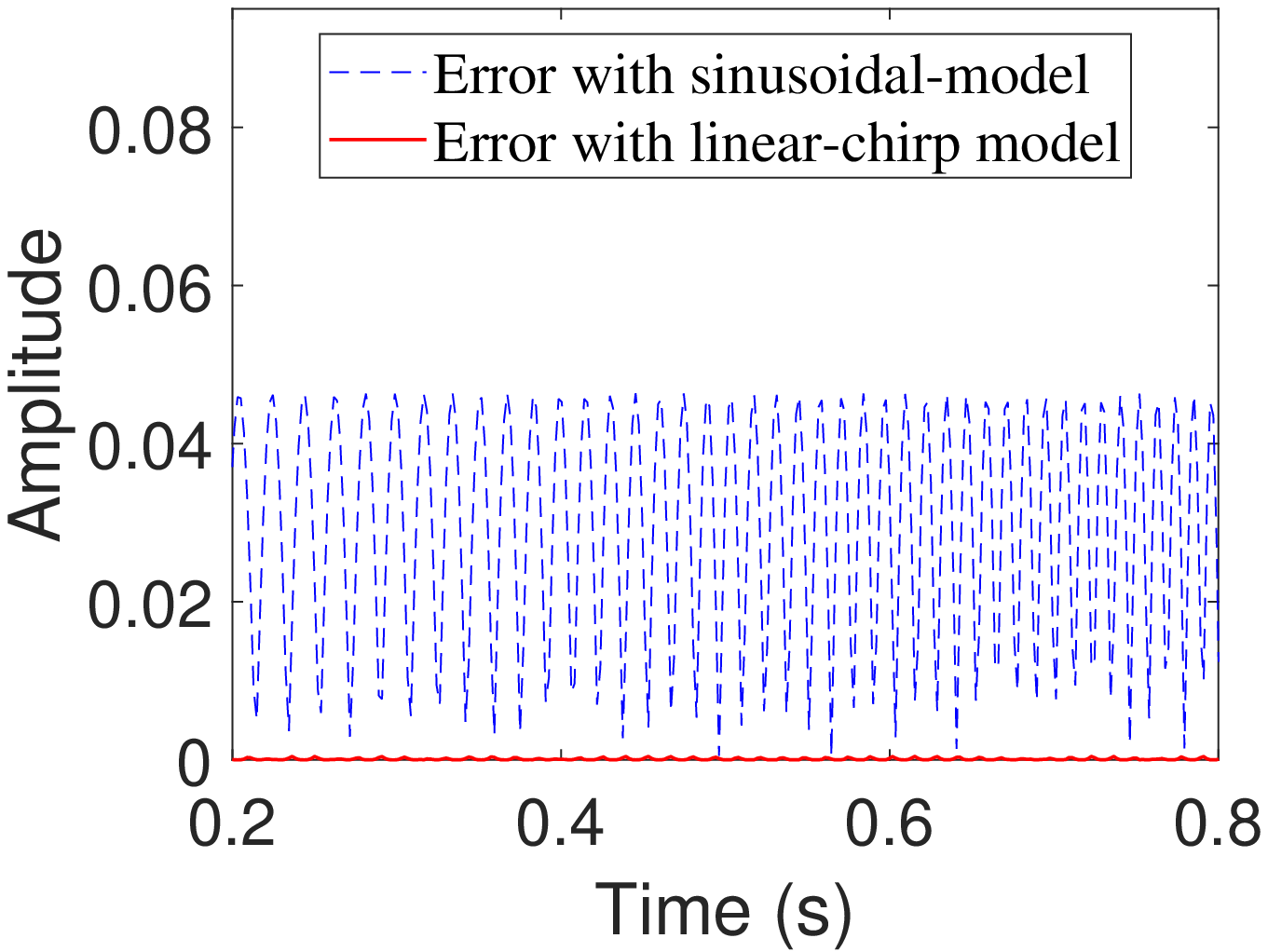}}
		\centerline{\fontsize{8.0pt}{\baselineskip}\selectfont (a)}
	\end{minipage}
	\hfill
	\begin{minipage}{0.48\linewidth}
		\centerline{\includegraphics[width=6.6cm]{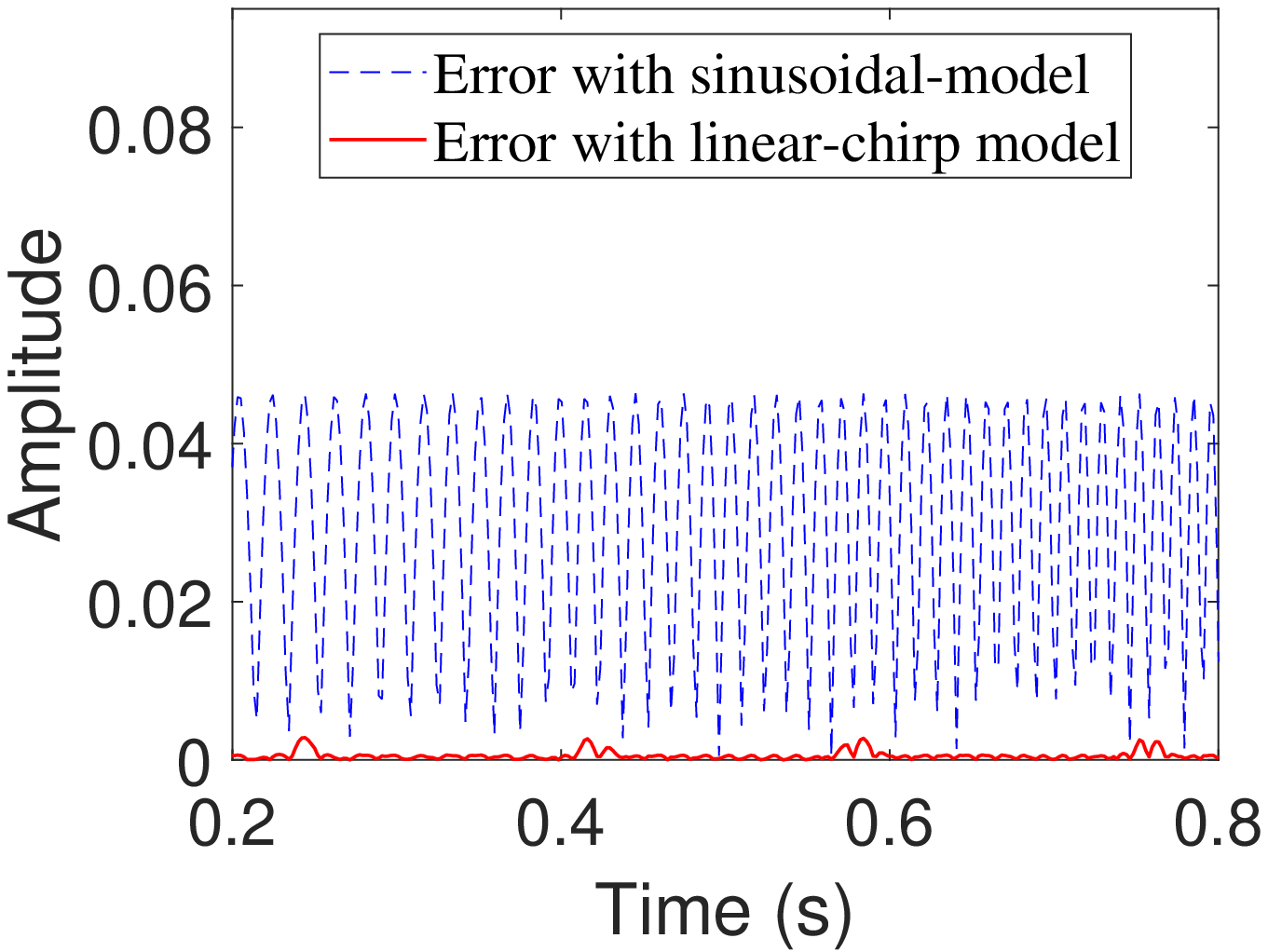}}
		\centerline{\fontsize{8.0pt}{\baselineskip}\selectfont (b)}
	\end{minipage}
	\caption { Comparison of linear chirp model and sinusoidal models. Left:  ground truth $\phi^{\gp\gp}_k(t)$ is used in  \eqref{func_recover2_k}; Right: 
	an estimate of  $\phi^{\gp\gp}_k(t)$ is used in  \eqref{func_recover2_k}.
	}
	\label{fig:mono-LFM-signal}
\end{figure}

\begin{figure}[th]
	\begin{minipage}{0.48\linewidth}
		\centerline{\includegraphics[width=6.6cm]{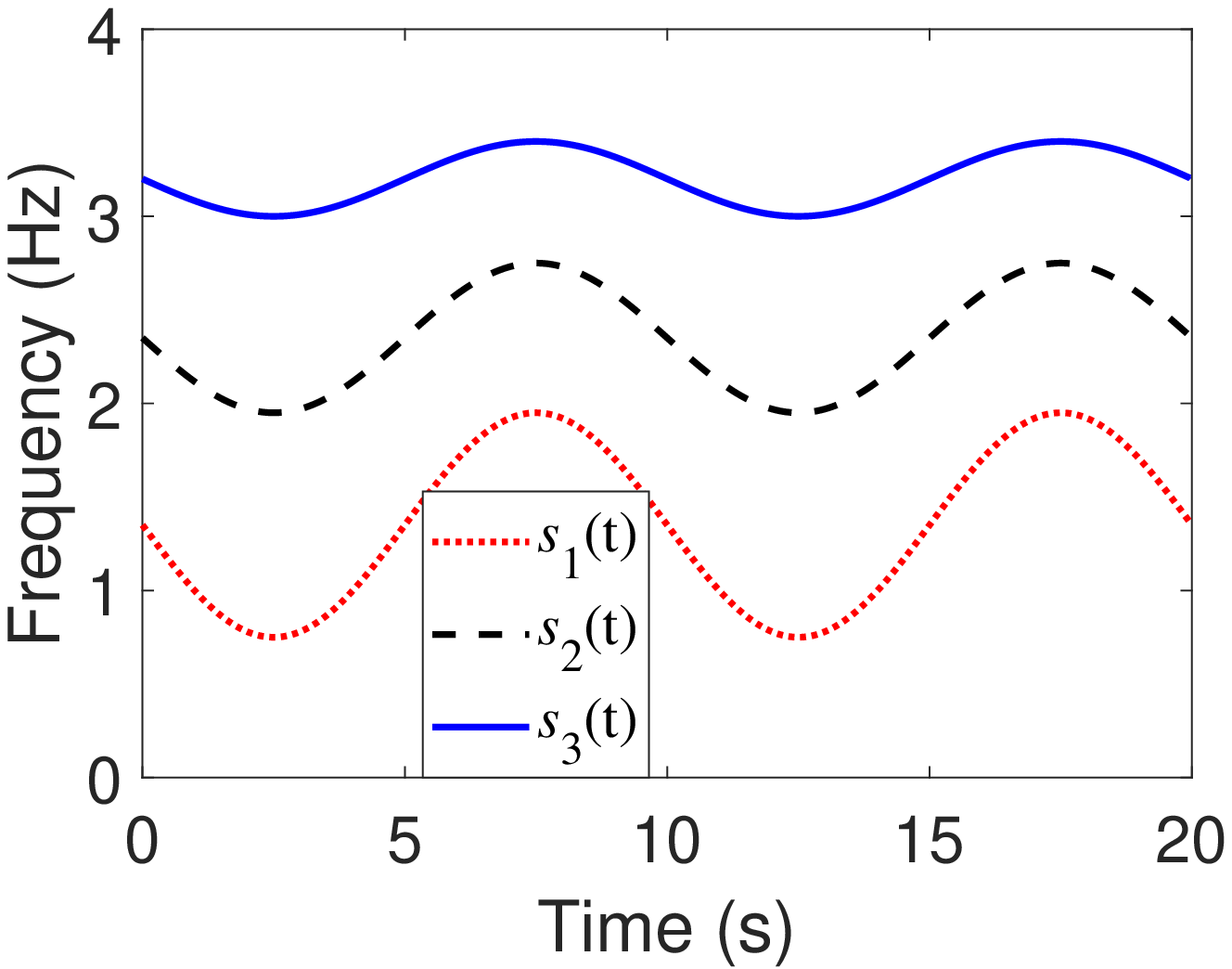}}
		\centerline{\fontsize{8.0pt}{\baselineskip}\selectfont (a)}
	\end{minipage}
	\hfill
	\begin{minipage}{0.48\linewidth}
		\centerline{\includegraphics[width=6.6cm]{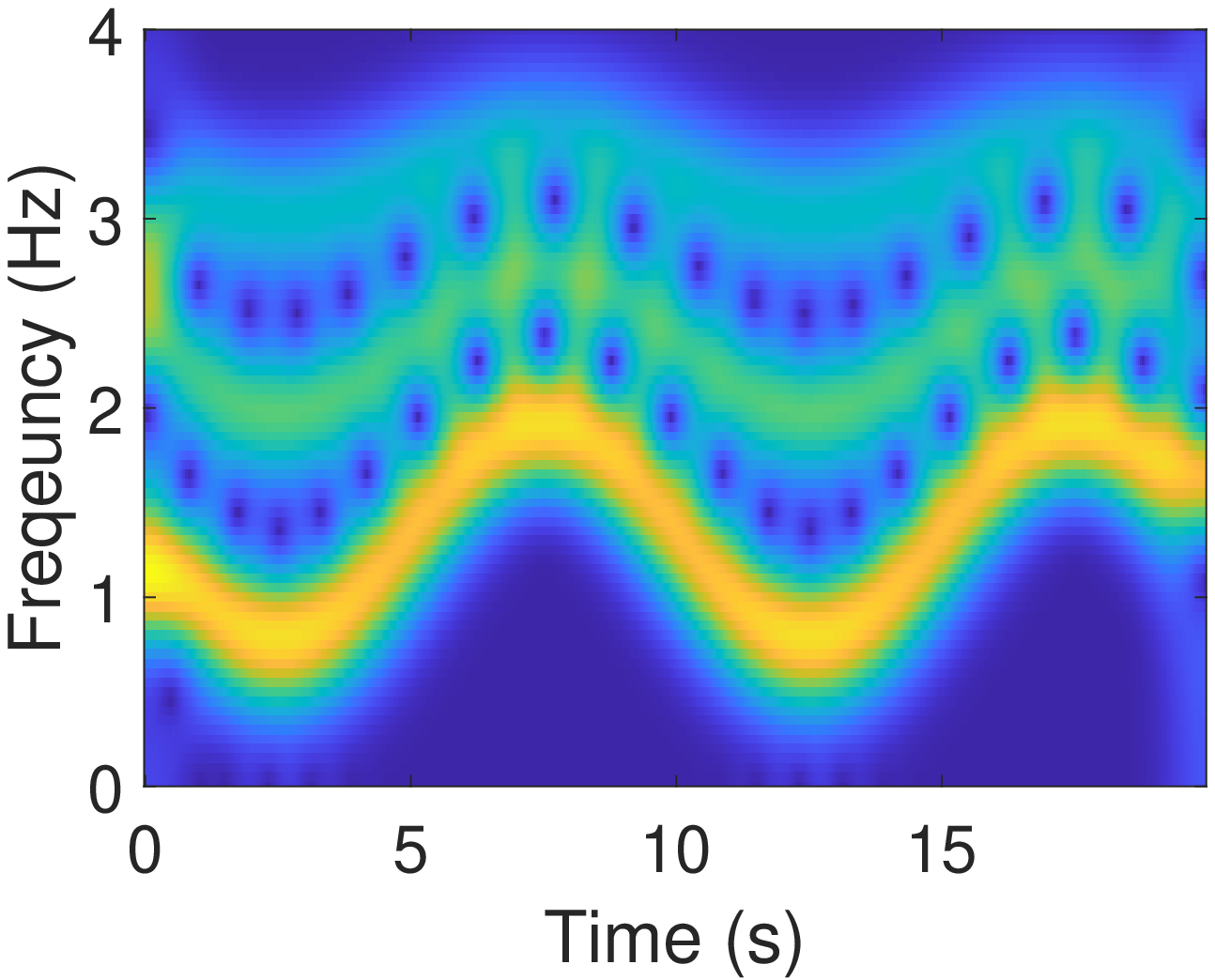}}
		\centerline{\fontsize{8.0pt}{\baselineskip}\selectfont (b)}
	\end{minipage}
	\caption {IFs (left panel) and STFT (right panel) of  $s(t)$ in \eqref{three_nonlinear_FM}.}
	\label{fig:the three-comp signal}
\end{figure}

Next we consider a three-component nonlinear FM signal,
\begin{equation}
\label{three_nonlinear_FM}
\begin{array}{l}
s(t)=s_1(t)+s_2(t)+s_3(t)
\\=\cos \left(2.7\pi t + 6\cos(0.2\pi t)\right)+ \frac 23\cos \left(4.7\pi t+4\cos(0.2\pi t)\right) \\+
\frac 12\cos \left(6.4\pi t+2\cos(0.2\pi t)\right),
\end{array}
\end{equation}
where $ t\in [0, 20]$. The number of sampling points is $N=512$, namely sampling rate is $F_s=25.6$ Hz.
The IFs of  $s_1(t)$, $s_2(t)$ and $s_3(t)$ are  $\phi'_1(t)=1.35-0.6\sin(0.2\pi t)$, $\phi'_2(t)=2.35-0.4\sin(0.2\pi t)$ and $\phi'_3(t)=3.2-0.2\sin(0.2\pi t)$, respectively.
The chirp rates of  $s_1(t)$, $s_2(t)$ and $s_3(t)$ are  $\phi''_1(t)=-0.12 \pi \cos(0.2\pi t)$, $\phi''_2(t)=-0.08 \pi \cos(0.2\pi t)$ and $\phi''_3(t)=-0.04 \pi \cos(0.2\pi t)$, respectively.
 Fig.\ref{fig:the three-comp signal} shows the IFs of the three components and STFT of $s(t)$.

\begin{figure}[th]
	\centering
	\begin{minipage}{0.48\linewidth}
		\centerline{\includegraphics[width=6.6cm]{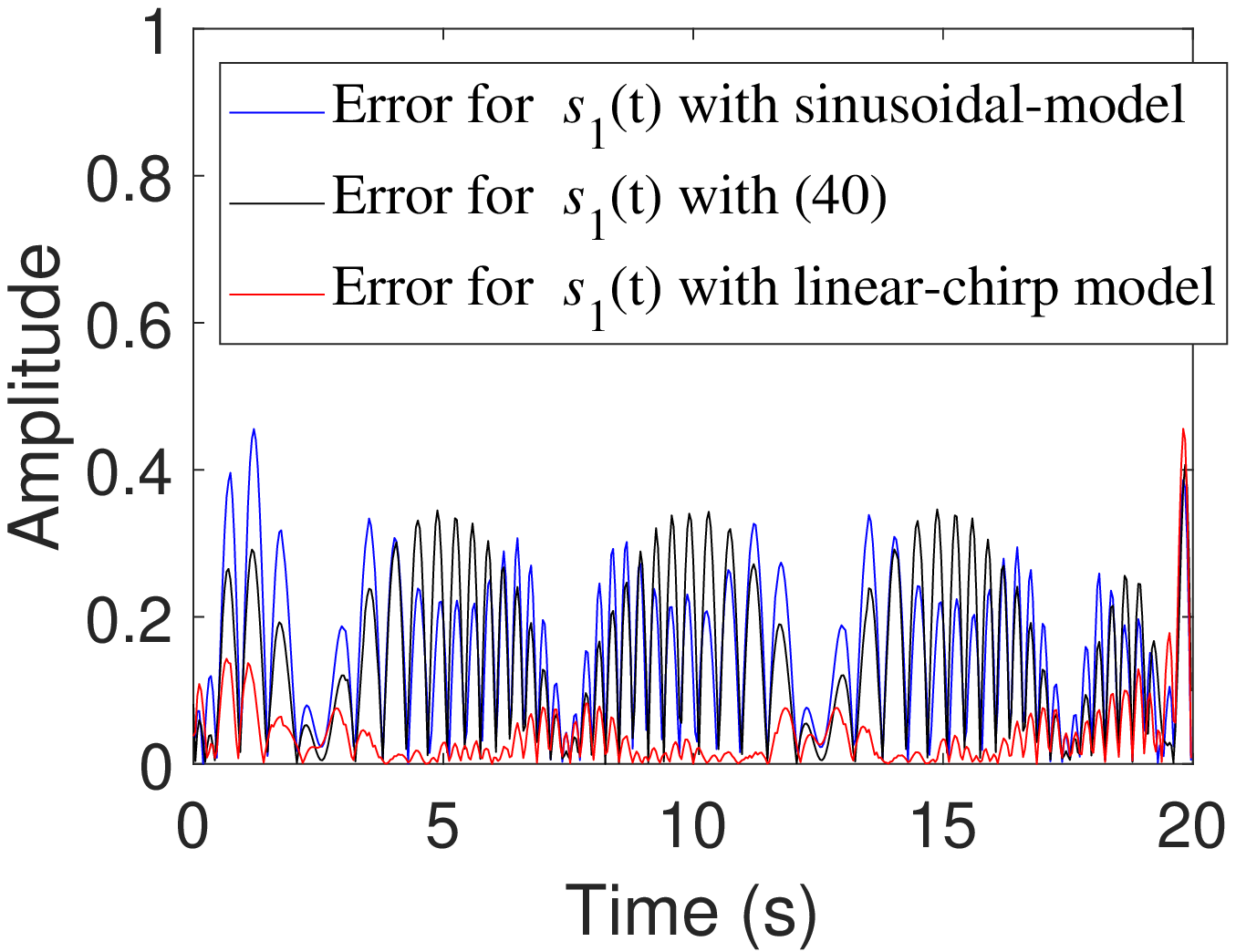}}
		\centerline{\fontsize{8.0pt}{\baselineskip}\selectfont (a)}
	\end{minipage}
	\hfill
	\begin{minipage}{0.48\linewidth}
		\centerline{\includegraphics[width=6.6cm]{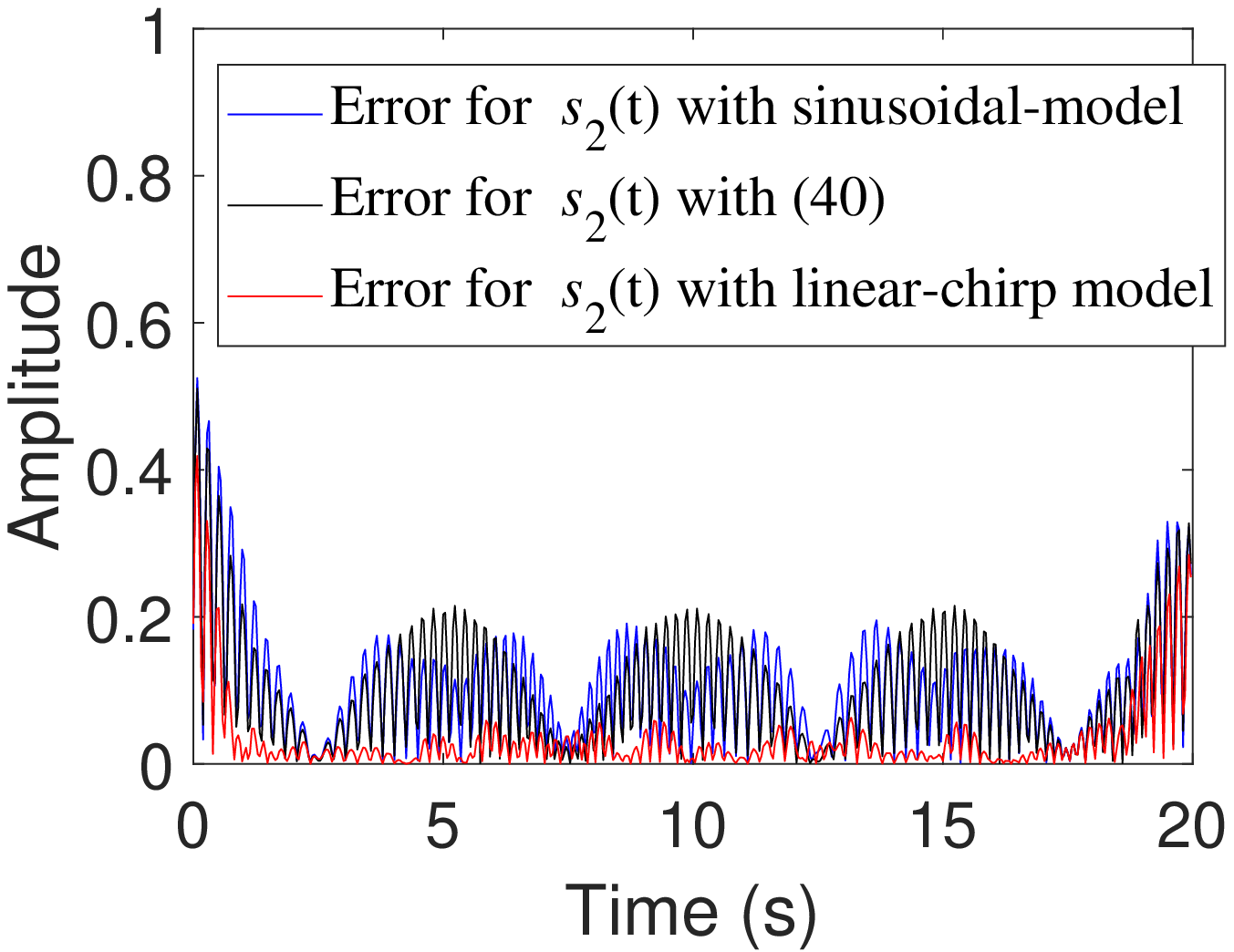}}
		\centerline{\fontsize{8.0pt}{\baselineskip}\selectfont (b)}
	\end{minipage}
	\begin{minipage}{0.48\linewidth}
	\centerline{\includegraphics[width=6.6cm]{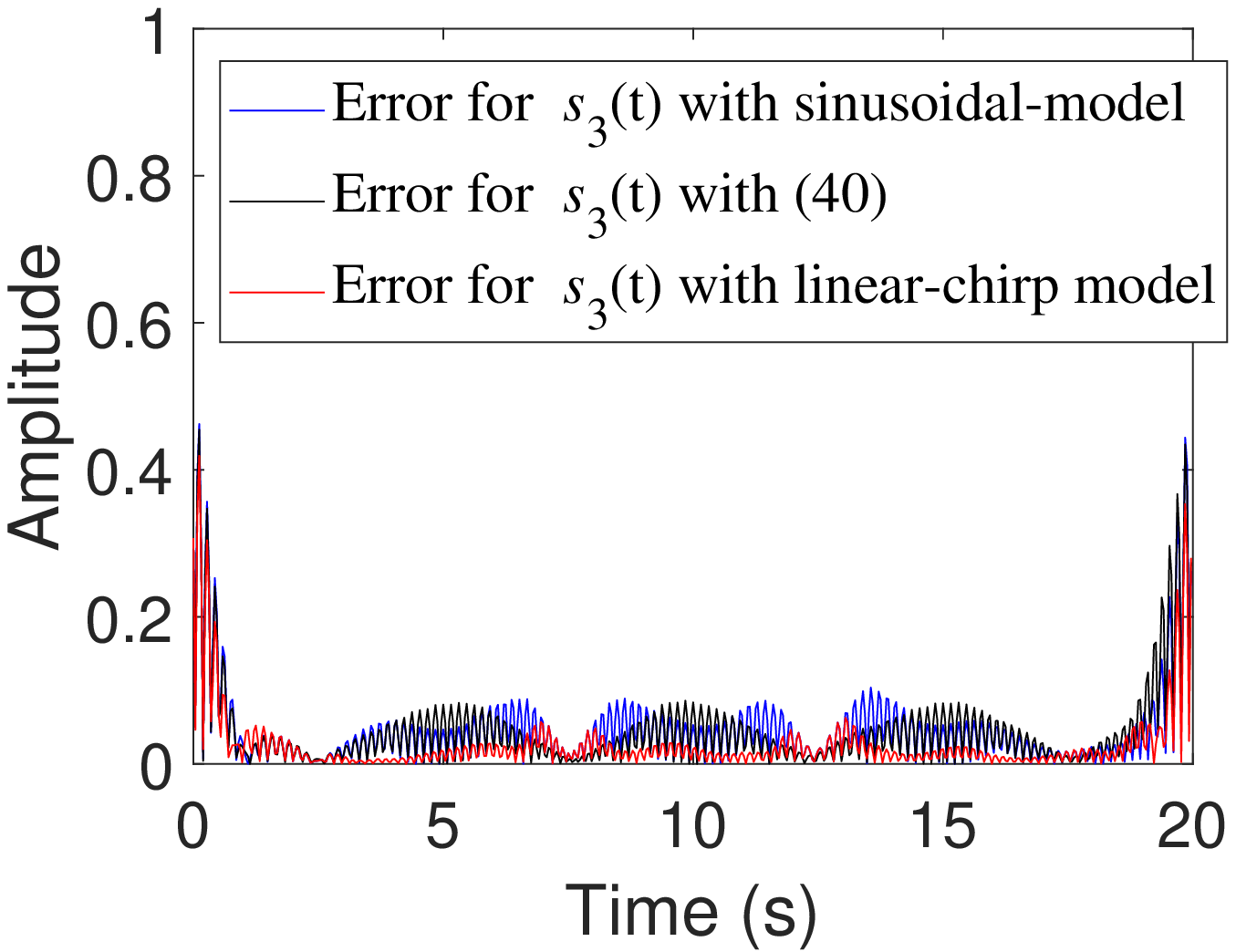}}
	\centerline{\fontsize{8.0pt}{\baselineskip}\selectfont (c)}
	\end{minipage}
	\hfill
	\begin{minipage}{0.48\linewidth}
	\centerline{\includegraphics[width=6.6cm]{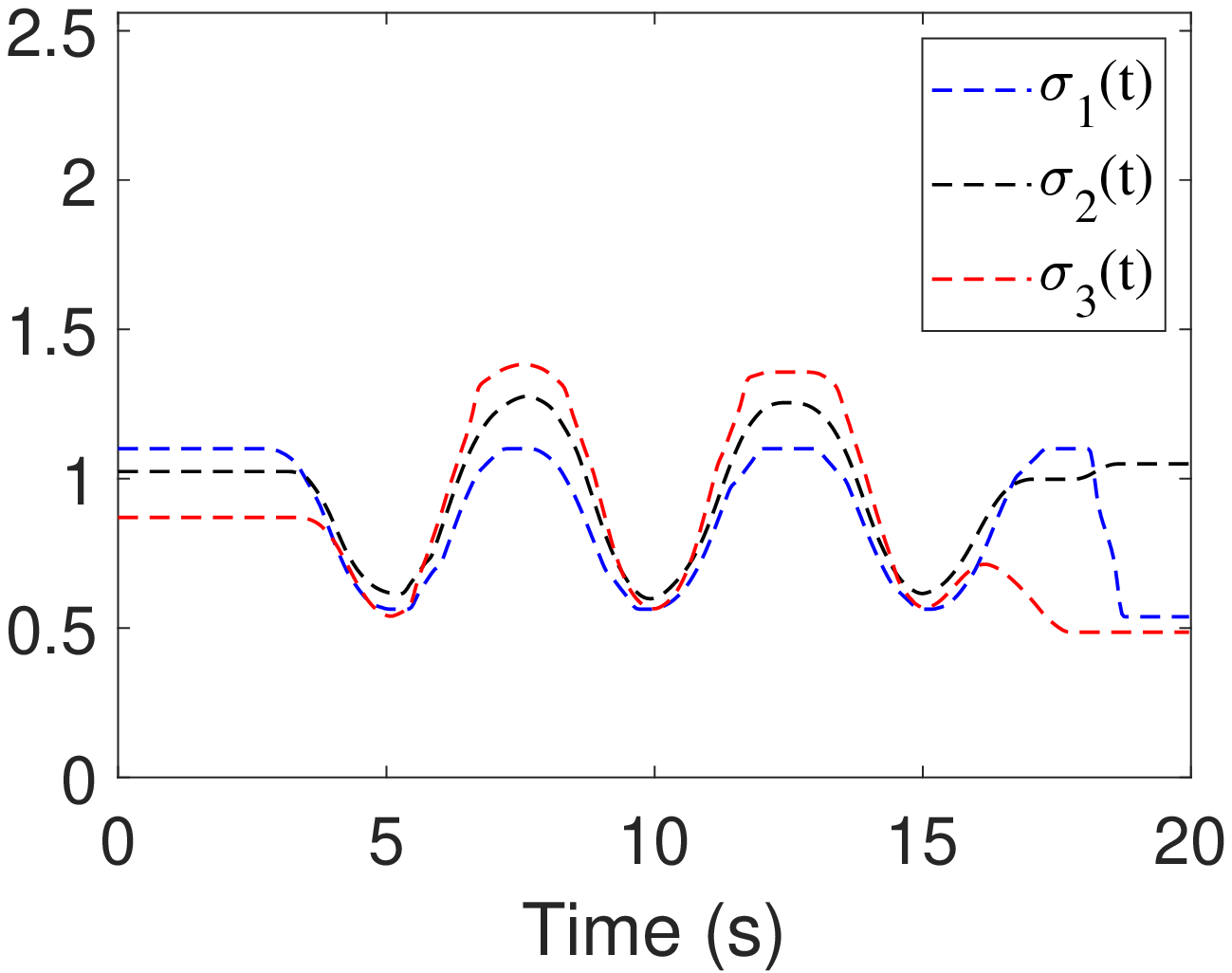}}
	\centerline{\fontsize{8.0pt}{\baselineskip}\selectfont (d)}
	\end{minipage}

	\caption {Comparison of our proposed ASSO scheme of linear chirp model (solid red lines) with other methods: the recovery errors of the three components $s_1(t)$ (a), $s_2(t)$ (b) and $s_3(t)$ (c) with different methods, and the estimated time-varying parameters $\gs_p(t)$ for the three components (d).}
	\label{fig:results of three-comp signal}
\end{figure}

Fig.\ref{fig:results of three-comp signal}(a), (b) and (c) show the absolute errors $\big| \wc s_k(t) -s_k(t) \big|, k = 1,2,3$ for  $s_1(t)$, $s_2(t)$ and $s_3(t)$, respectively. We compare the three 
 methods here, namely the ASSO with sinusoidal signal-based model (sinusoidal-model) in \eqref{func_recover}, the recovery equation \eqref{recovery_their} derived from \cite{LM1} and our proposed ASSO with linear-chirp model in \eqref{ASSO_recon_generalized}. 
Based on Algorithm 1 and 2, we recover the three components one by one. Here for both ASSO methods with sinusoidal-model and linear-chirp model, we update the time-varying window parameters $\gs_{\rm R}(t)$ and $\gs_p(t)$ for each component. 
For the method in \cite{LM1}, namely \eqref{recovery_their}, the IF estimation $\wh \go_f^{[2]}(t, \varphi_k(t))$ and the chirp rate estimation $\wh q_f\big(t, \varphi_k(t)\big)$ are obtained as those in \cite{LM1} with $\varphi_k(t)$ as the IF estimation from the 2nd-order SST.

Meanwhile, since \cite{LM1} considers constant window, for the sake of fairness, we also update the window parameter $\gs$ for each component as the average of $\gs_p(t)$ used for ASSO, see Fig.\ref{fig:results of three-comp signal}(d). Obviously, the recovery error of  our ASSO of linear-chirp model is less than the other two methods for all the three components. Due to the boundary effect, the recovery errors at the two ends of the signal are larger than the center. 
  It is worth to note that the proposed algorithm in this paper is a localized method which is suitable to process the real-world signal with consecutive input efficiently.    

\begin{figure}[th]
	\centering
	\begin{minipage}{0.48\linewidth}
		\centerline{\includegraphics[width=6.4cm]{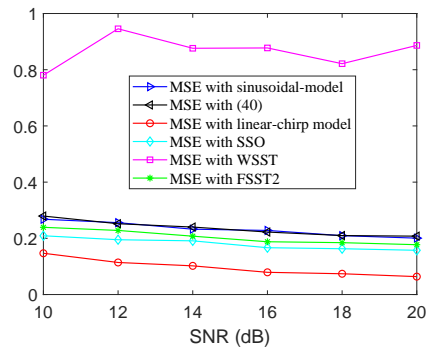}}
	\end{minipage}
	
	\caption {Comparison of our proposed ASSO scheme of linear chirp model (red circles) with other methods: recovery MSE under different SNRs.}
	\label{fig:MSE-SNR}
\end{figure}

Next we consider the performance of our proposed method in a noisy  environment. We add Gaussian noises to the source signal $s(t)$ in \eqref{three_nonlinear_FM} with signal-to-noise ratio (SNR) ranging from 10dB to 20dB. We use the mean-square error (MSE) to evaluate the recovery performance, which is defined by
\begin{equation}
\label{def_MSE}
{\rm MSE}_s = \frac 1K \sum_{k=1}^{K} \frac{\|s_k(t)- \wc s_k (t)\|_2} {\|s_k(t)\|_2}, 
\end{equation} 
where $ \wc s_k (t)$ is the recovery function of $s_k(t)$.
We also perform the conventional SSO in \cite{Chui_M15},  CWT-based SST (WSST) in \cite{Daub_Lu_Wu11} and the second-order STFT-based SST (FSST2) in \cite{MOM15} for comparison.
Fig.\ref{fig:MSE-SNR} demonstrates that the proposed ASSO scheme of linear-chirp model is more accurate than the other methods for component recovery. Because of the bound effect mentioned above, here we just consider the center part of the signal, namely $t\in [2.5, 17.5]$, to calculate the MSE in \eqref{def_MSE}.

\clearpage

	\subsection {Radar echoes}
	
\begin{figure}[tbh]
	\centering
	\begin{minipage}{0.48\linewidth}
		\centerline{\includegraphics[width=6.6cm]{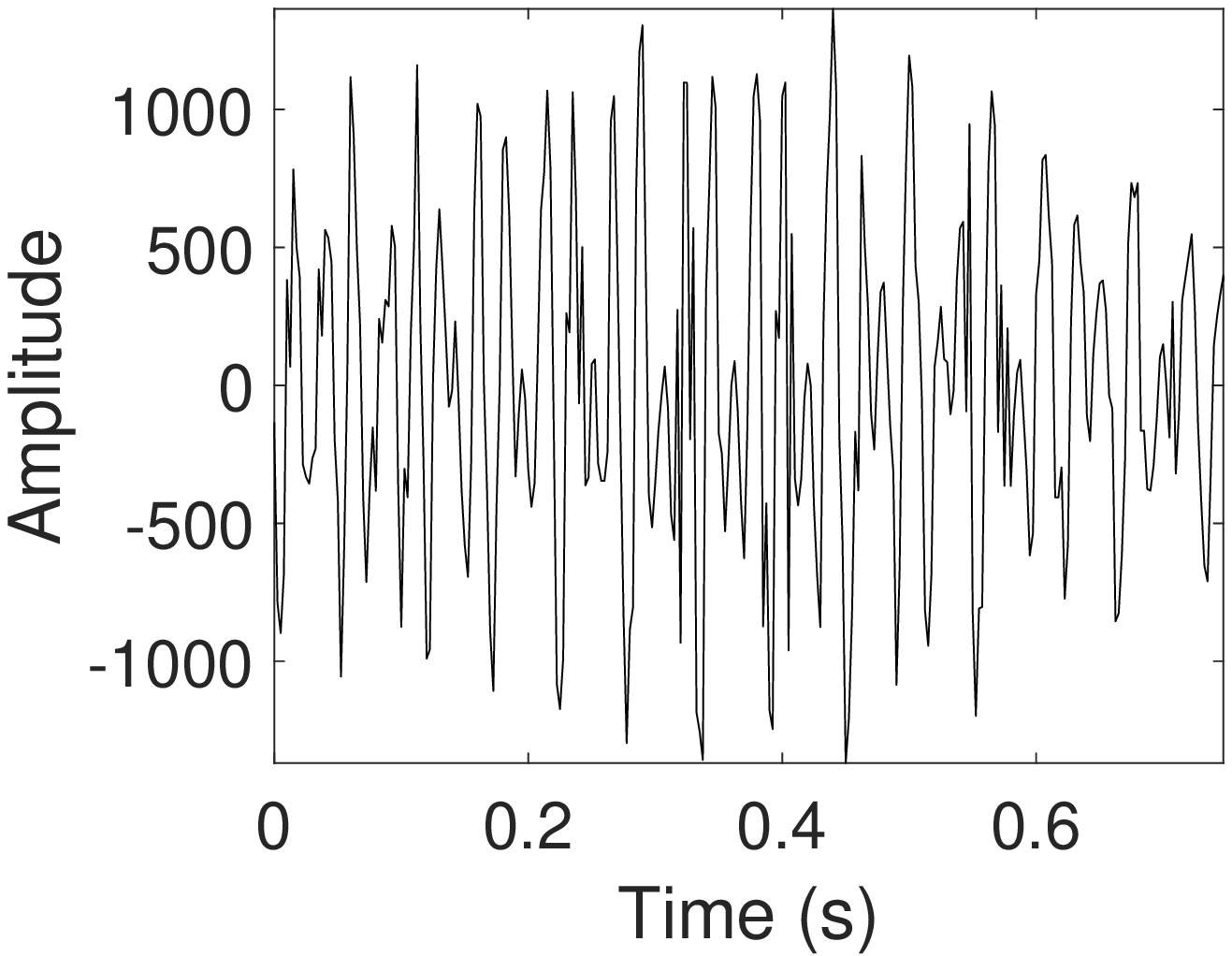}}
		\centerline{\fontsize{8.0pt}{\baselineskip}\selectfont (a)}
	\end{minipage}
	\hfill
	\begin{minipage}{0.48\linewidth}
		\centerline{\includegraphics[width=6.6cm]{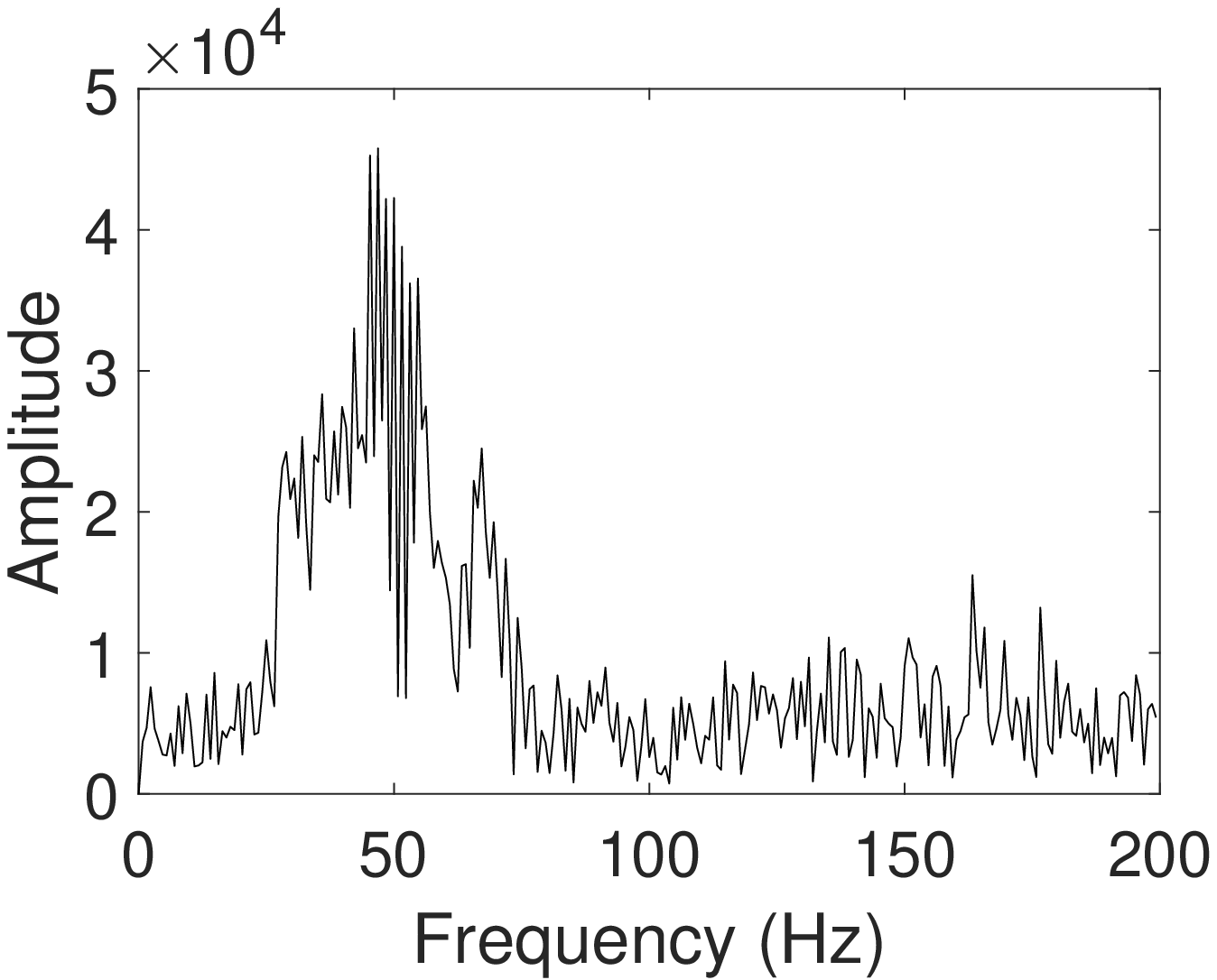}}
		\centerline{\fontsize{8.0pt}{\baselineskip}\selectfont (b)}
	\end{minipage}
	\begin{minipage}{0.48\linewidth}
		\centerline{\includegraphics[width=6.6cm]{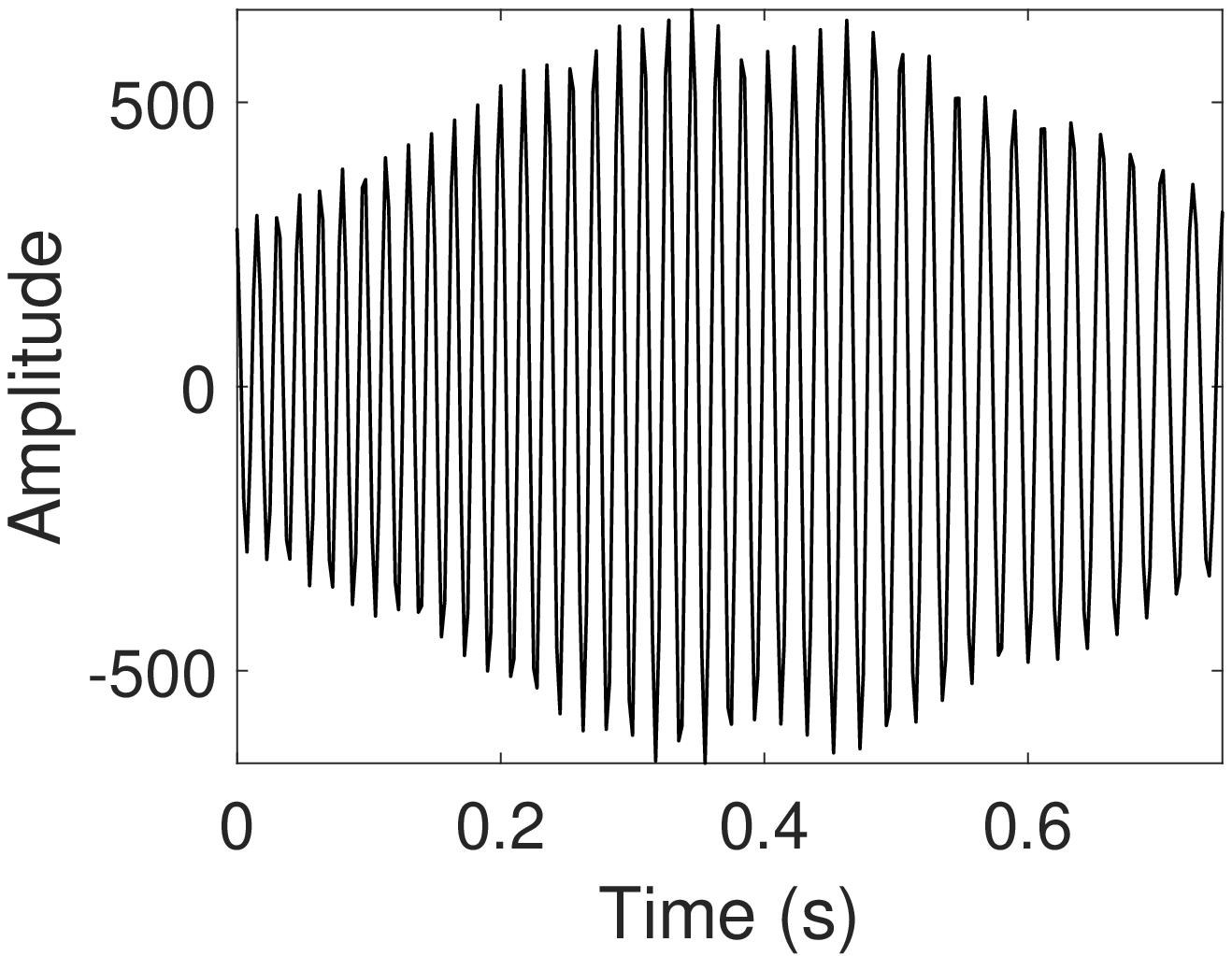}}
		\centerline{\fontsize{8.0pt}{\baselineskip}\selectfont (c)}
	\end{minipage}
	\hfill
	\begin{minipage}{0.48\linewidth}
		\centerline{\includegraphics[width=6.6cm]{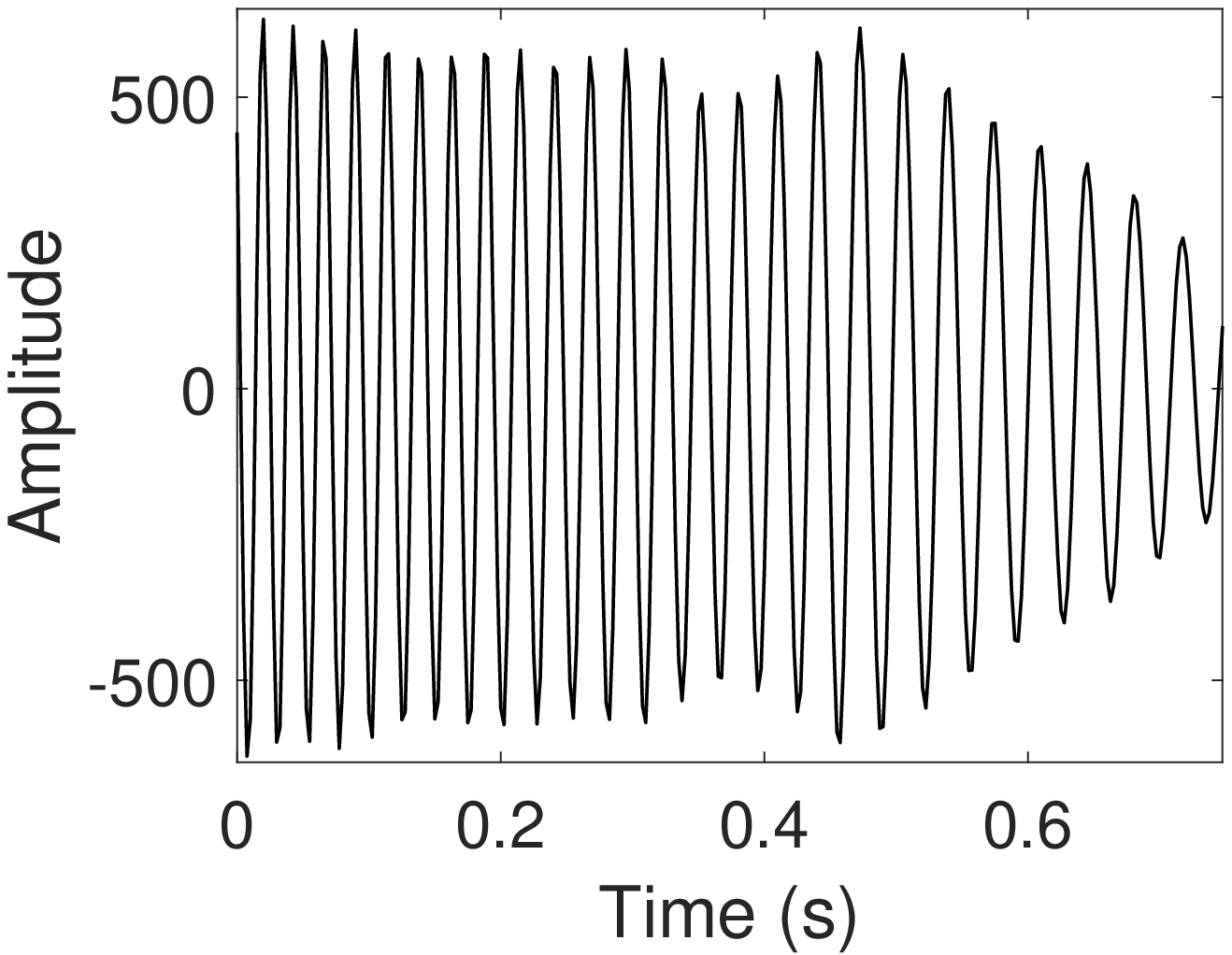}}
		\centerline{\fontsize{8.0pt}{\baselineskip}\selectfont (d)}
	\end{minipage}

\begin{minipage}{0.48\linewidth}
	\centerline{\includegraphics[width=6.6cm]{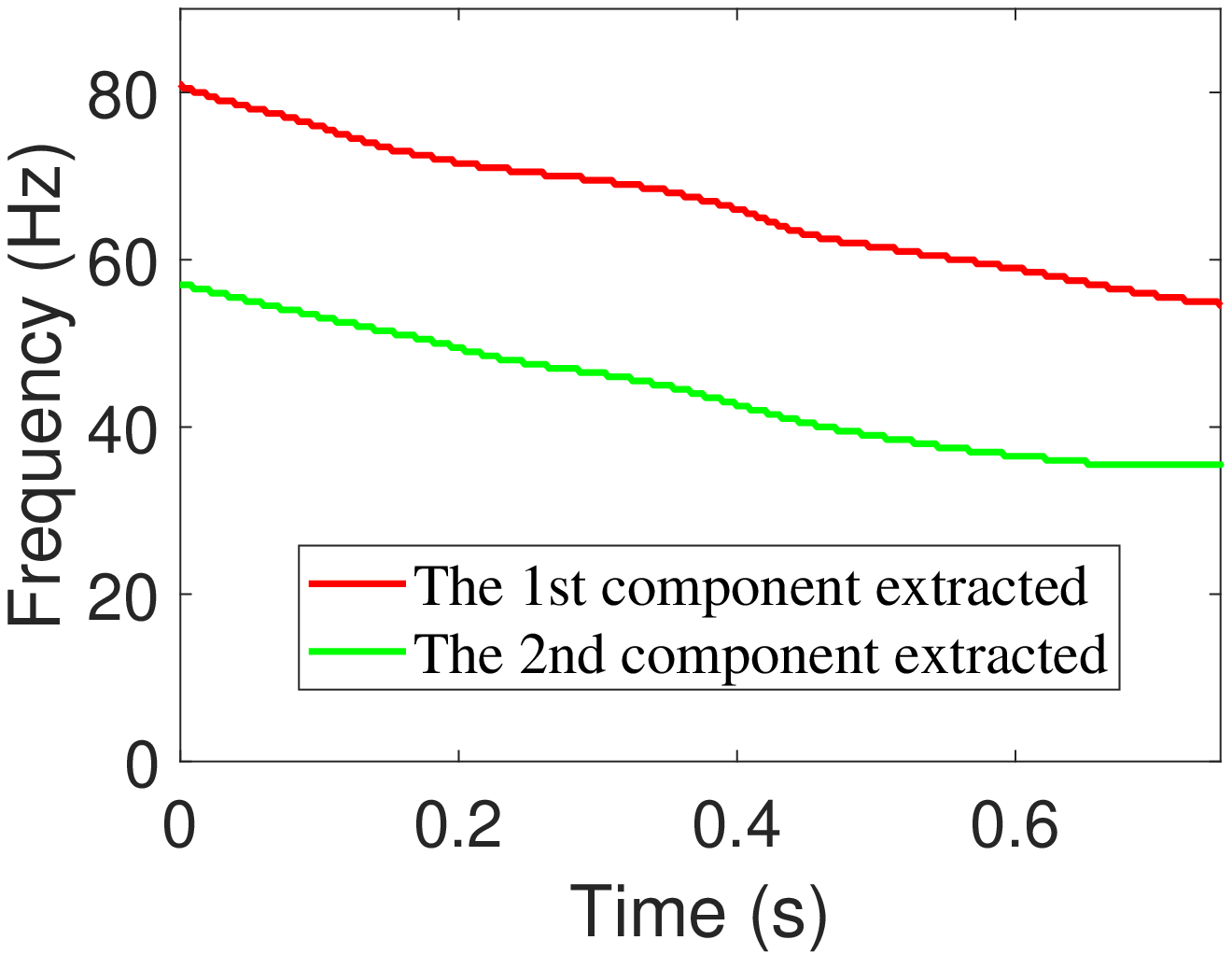}}
	\centerline{\fontsize{8.0pt}{\baselineskip}\selectfont (e)}
\end{minipage}
\hfill
\begin{minipage}{0.48\linewidth}
	\centerline{\includegraphics[width=6.6cm]{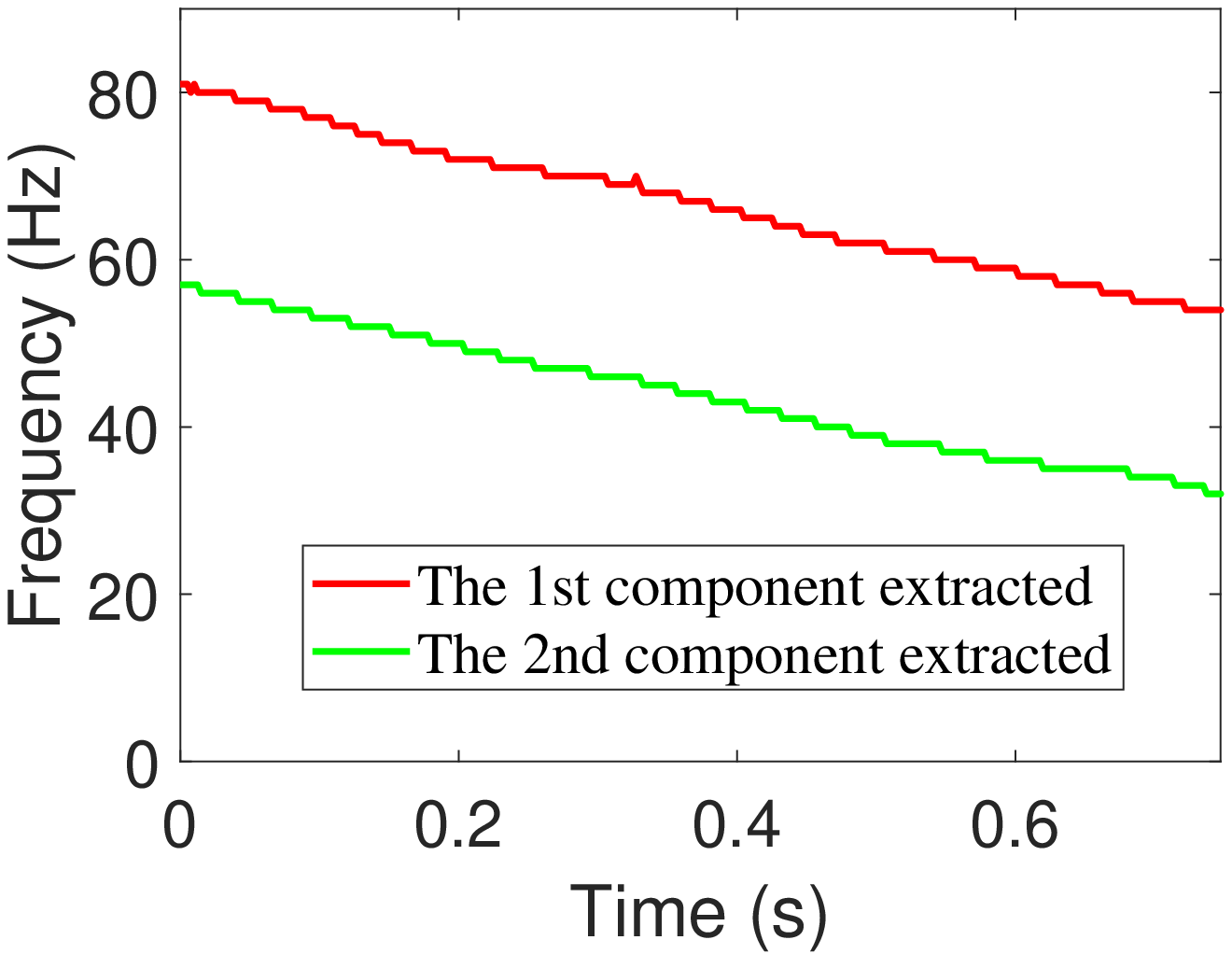}}
	\centerline{\fontsize{8.0pt}{\baselineskip}\selectfont (f)}
\end{minipage}
	
	\caption {Radar data and its estimation results: source data (a), spectrum (b), recovered Component 1 (c), recovered Component 2 (d) 
	IF estimated by ASSO (e) and  IF estimated by FSST2 (f).}
	\label{fig:radar_data}
\end{figure}

Here we consider the real data from a conventional low-resolution very-high-frequency (VHF) surveillance radar. The pulse-repetition frequency (PRF) of the radar is 400 Hz.
Fig.\ref{fig:radar_data}(a) shows a truncation of the data to be processed, which consists of 300 discrete samples (we have more data outside this truncation). Note that these samples are corresponding to each radar echoes, which means the samples are obtained after the radar signal processing of matched filtering.  
Therefore the sampling rate here is equal to the PRF, namely 400Hz. 
Fig.\ref{fig:radar_data}(c) and (d) shows the recovered waveforms by the proposed ASSO. 
Fig.\ref{fig:radar_data}(e) and (f) show the IF estimation results by ASSO and FSST2, respectively.

Observe that there are two signal components.
Actually, the data here is collected when two targets in formation fly past the radar. 
The IFs are corresponding to the Doppler frequencies aroused the changes of targets' aspects and speeds with respect to the radar. Since the two targets in formation are close, they are located at the same range unit of the conventional low-resolution radar. It is hardly to distinguish the two targets and estimate their Doppler frequencies by conventional Fourier analysis, see Fig.\ref{fig:radar_data}(b) for the spectrum. 
It shows the Doppler frequency of aerial target varies approximate linearly and smoothly in \cite{Li_Ji06}. To be fair, the window parameters $\gs$ for FSST2 is equal to the average value of the global time-varying parameter $\sigma_{\rm R}(t)$ estimated in Step 2 of Algorithm 1. All the IFs in  
Fig.\ref{fig:radar_data} are estimated by the ridges directly (without filtering or curve fitting). Observe that the results of our method are much smoother than those of FSST2.

	\begin{figure}[th]
		\centering
		\begin{minipage}{0.48\linewidth}
			\centerline{\includegraphics[width=6.6cm]{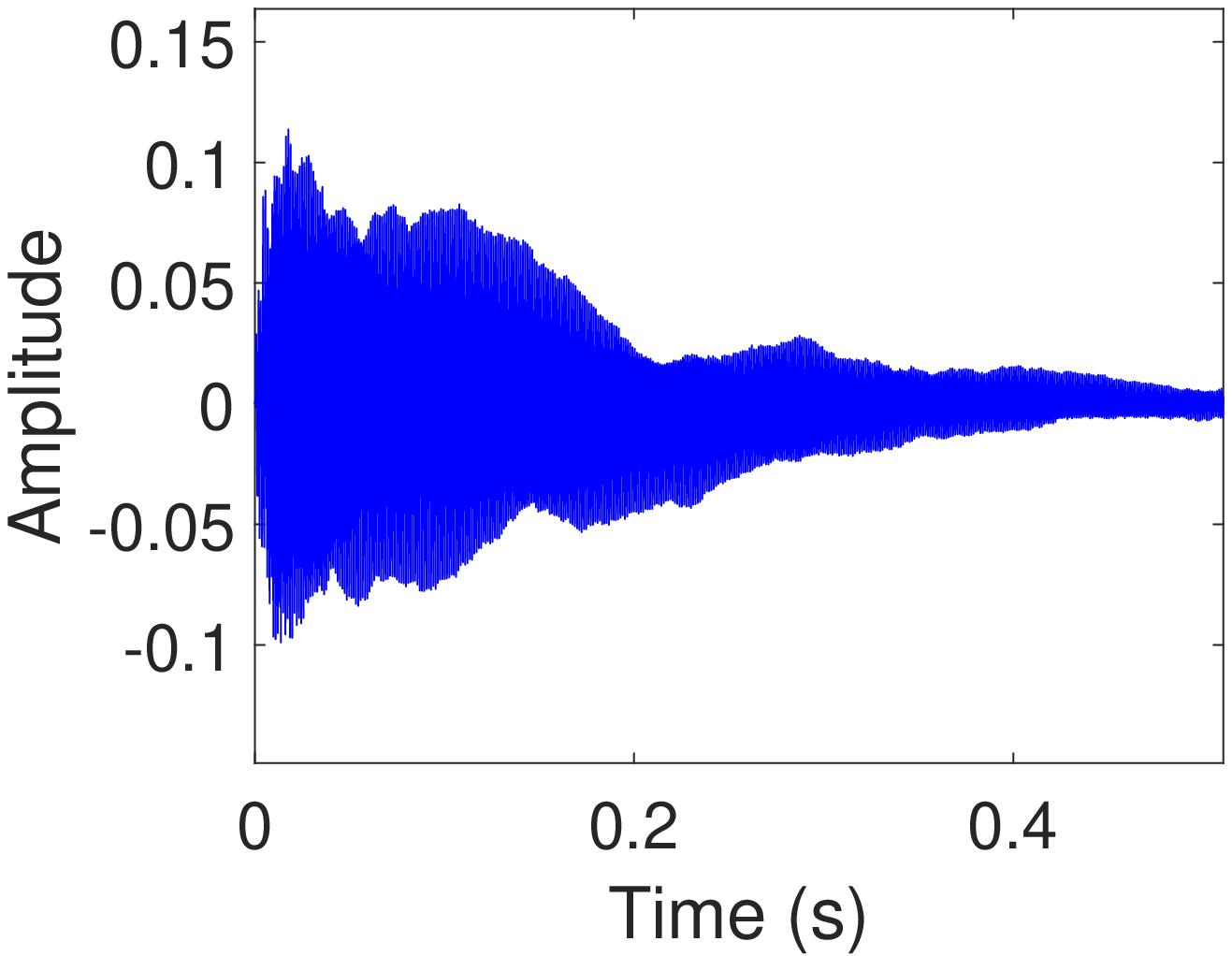}}
			\centerline{\fontsize{8.0pt}{\baselineskip}\selectfont (a)}
		\end{minipage}
		\hfill
		\begin{minipage}{0.48\linewidth}
			\centerline{\includegraphics[width=6.6cm]{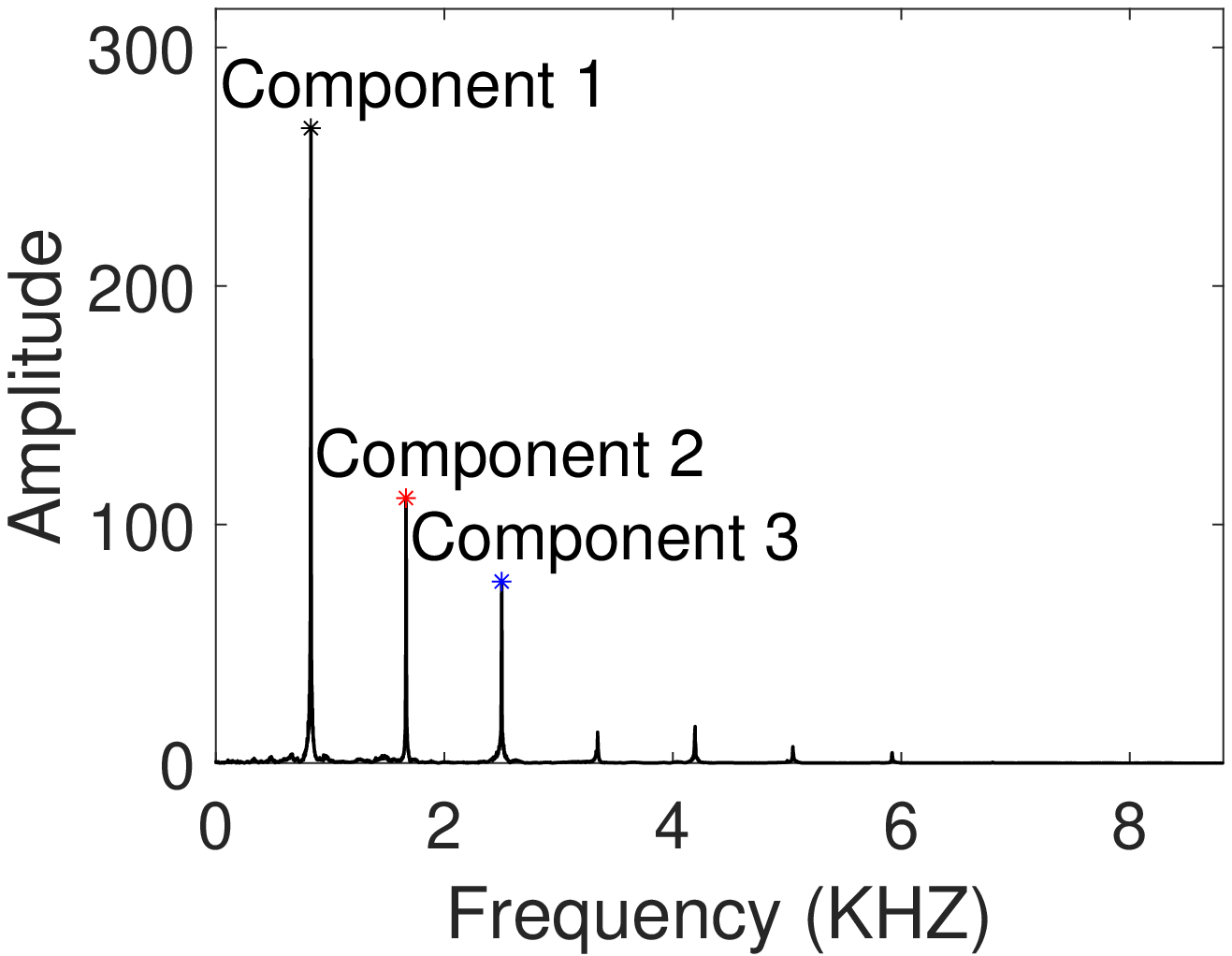}}
			\centerline{\fontsize{8.0pt}{\baselineskip}\selectfont (b)}
		\end{minipage}
		\begin{minipage}{0.48\linewidth}
			\centerline{\includegraphics[width=6.6cm]{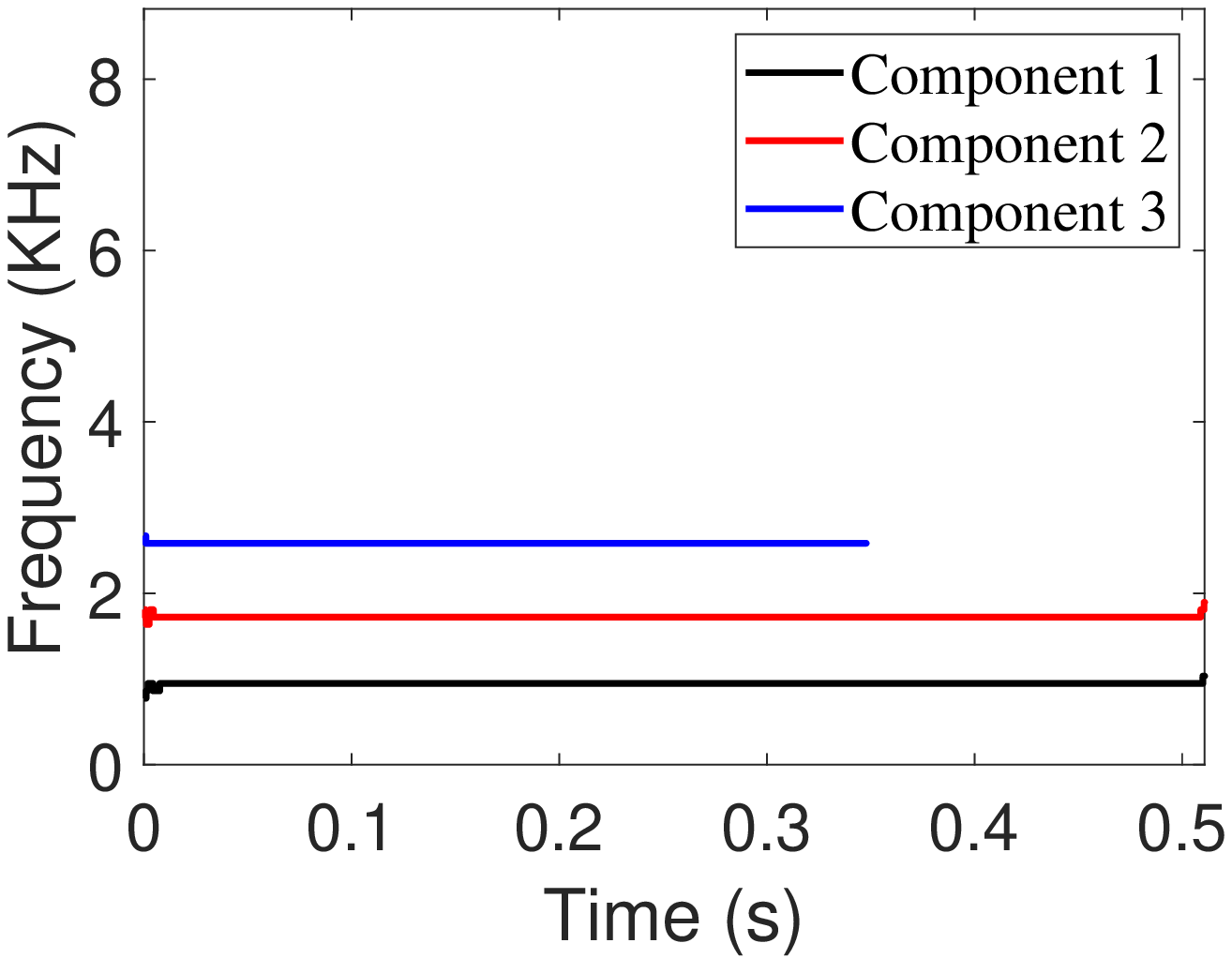}}
			\centerline{\fontsize{8.0pt}{\baselineskip}\selectfont (c)}
		\end{minipage}
		\hfill
		\begin{minipage}{0.48\linewidth}
			\centerline{\includegraphics[width=6.6cm]{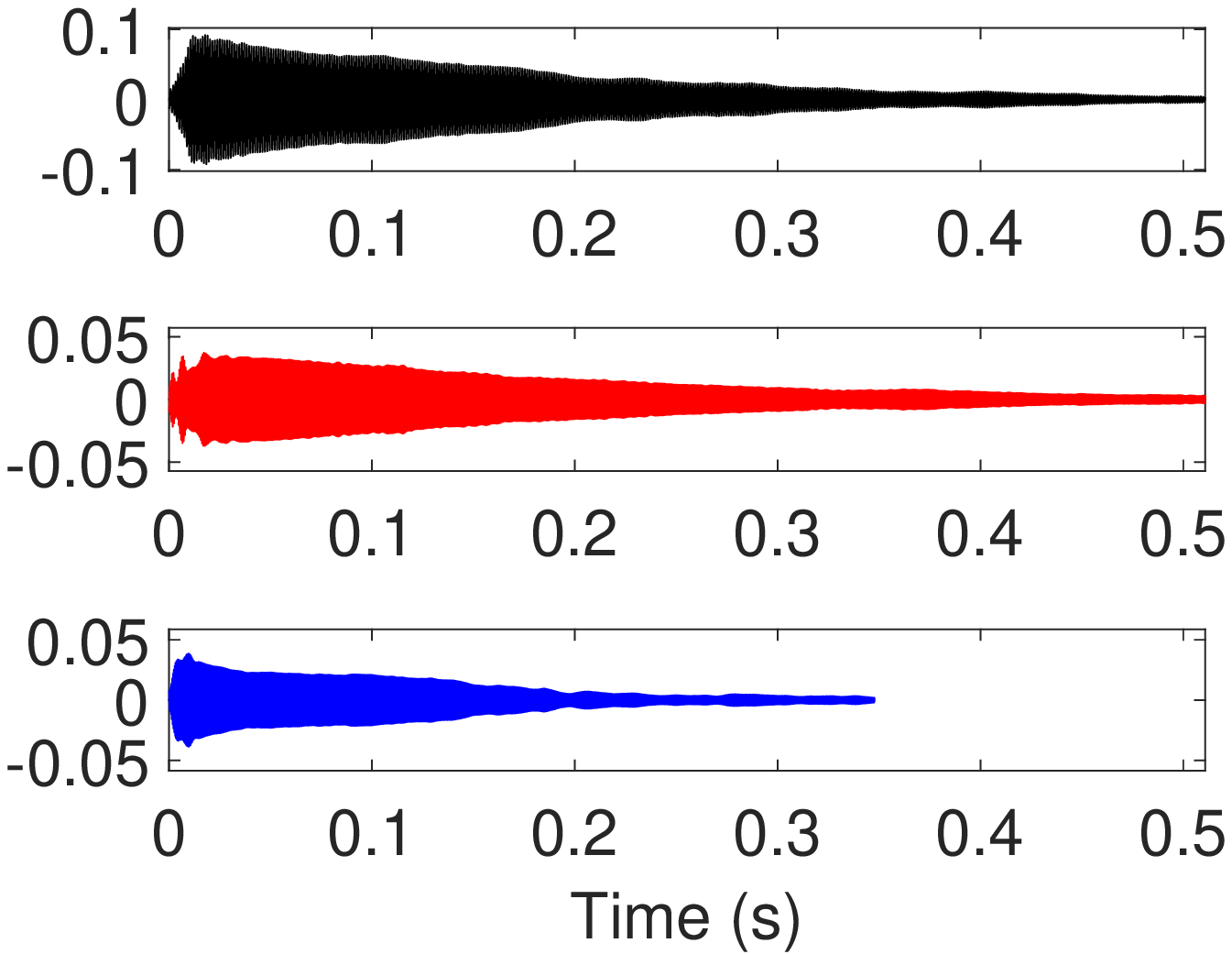}}
			\centerline{\fontsize{8.0pt}{\baselineskip}\selectfont (d)}
		\end{minipage}
		
		\caption {Music signal and its estimation results: source data (a), spectrum (b), IF estimation (c) and waveform reconstruction (d).}
		\label{fig:music_data}
	\end{figure}

\subsection{Audio signals}
	First we consider a music signal with sampling rate 44.1 KHz, see Fig.\ref{fig:music_data}(a) and (b) for the waveform and spectrum of the signal, respectively. 
	Note that there are 22,528 samples here, which is processed with a localized truncated Gaussian window with length of 256 points.
	Observe that this signal is a fundamental tone of music with three main harmonics, marked as Component 1, 2 and 3.
	The IFs of the three components are near to constant.
	Fig.\ref{fig:music_data}(c) and (d) show the estimated IFs and the recovered waveforms, respectively. The results in Fig.\ref{fig:music_data}(c) and (d) are corresponding to the three frequencies in (b).

\begin{figure}[th]
		\centering
		\begin{minipage}{0.48\linewidth}
			\centerline{\includegraphics[width=6.6cm]{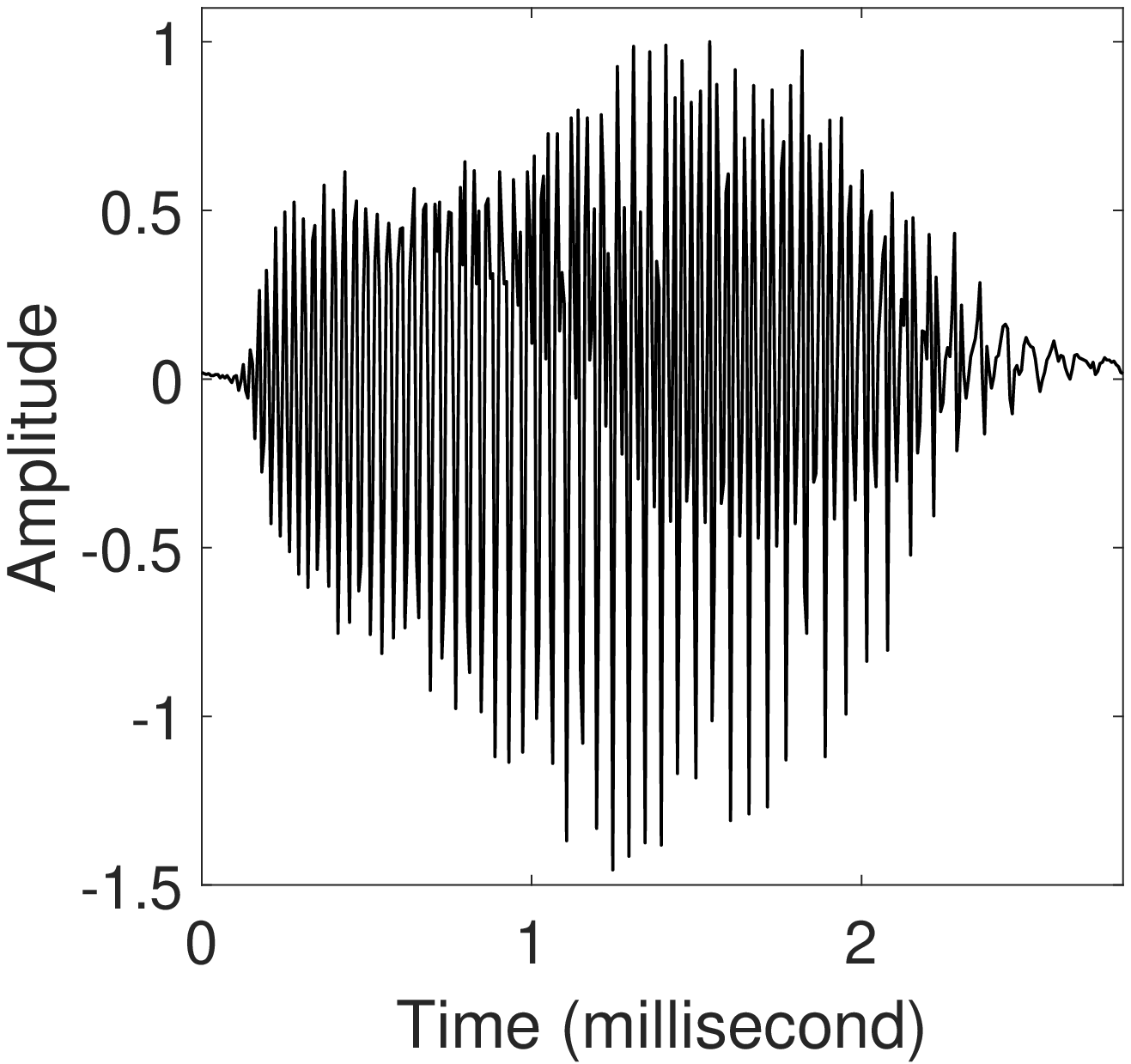}}
			\centerline{\fontsize{8.0pt}{\baselineskip}\selectfont (a)}
		\end{minipage}
		\hfill
		\begin{minipage}{0.48\linewidth}
			\centerline{\includegraphics[width=6.6cm]{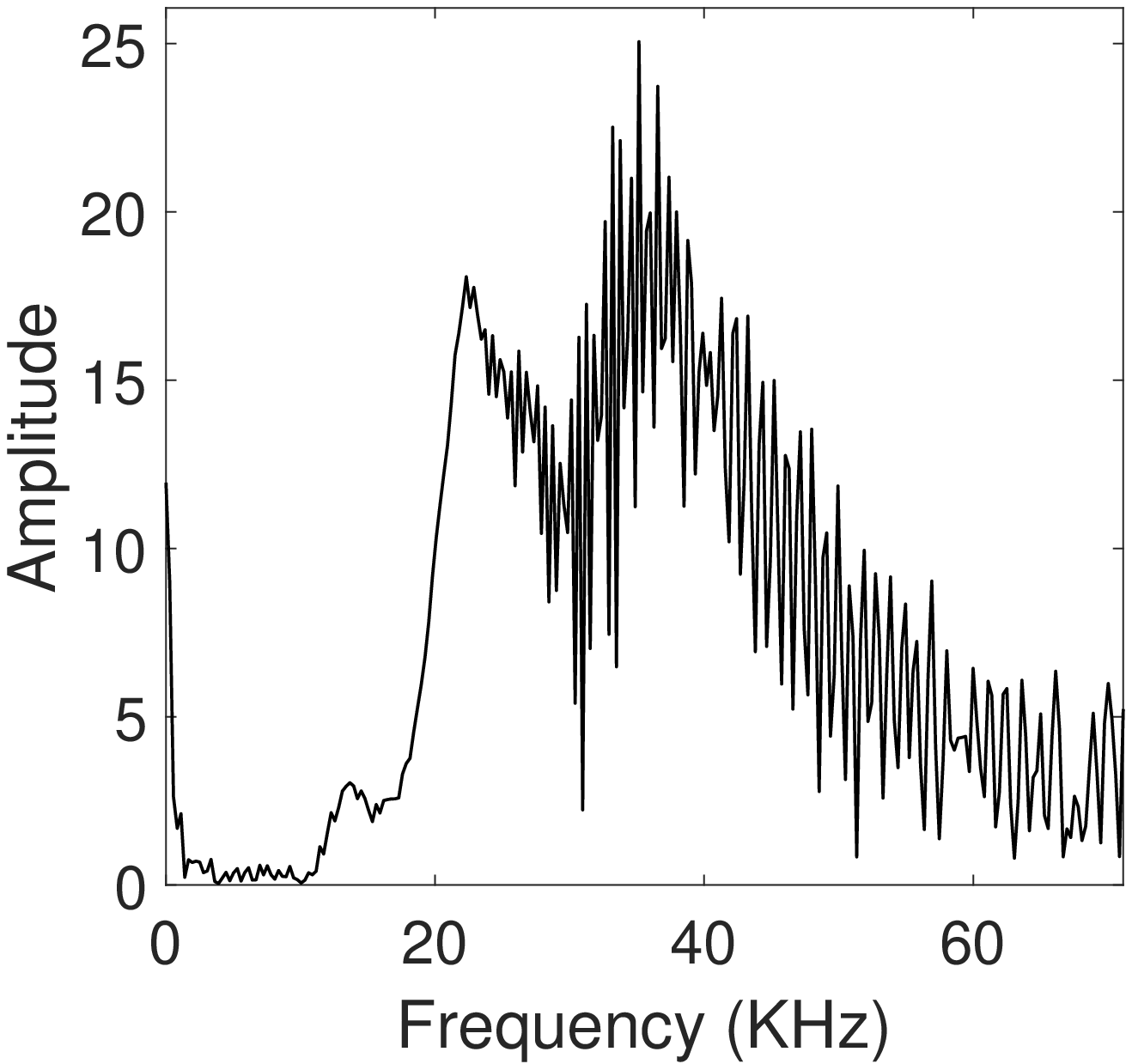}}
			\centerline{\fontsize{8.0pt}{\baselineskip}\selectfont (b)}
		\end{minipage}
		\begin{minipage}{0.48\linewidth}
			\centerline{\includegraphics[width=6.6cm]{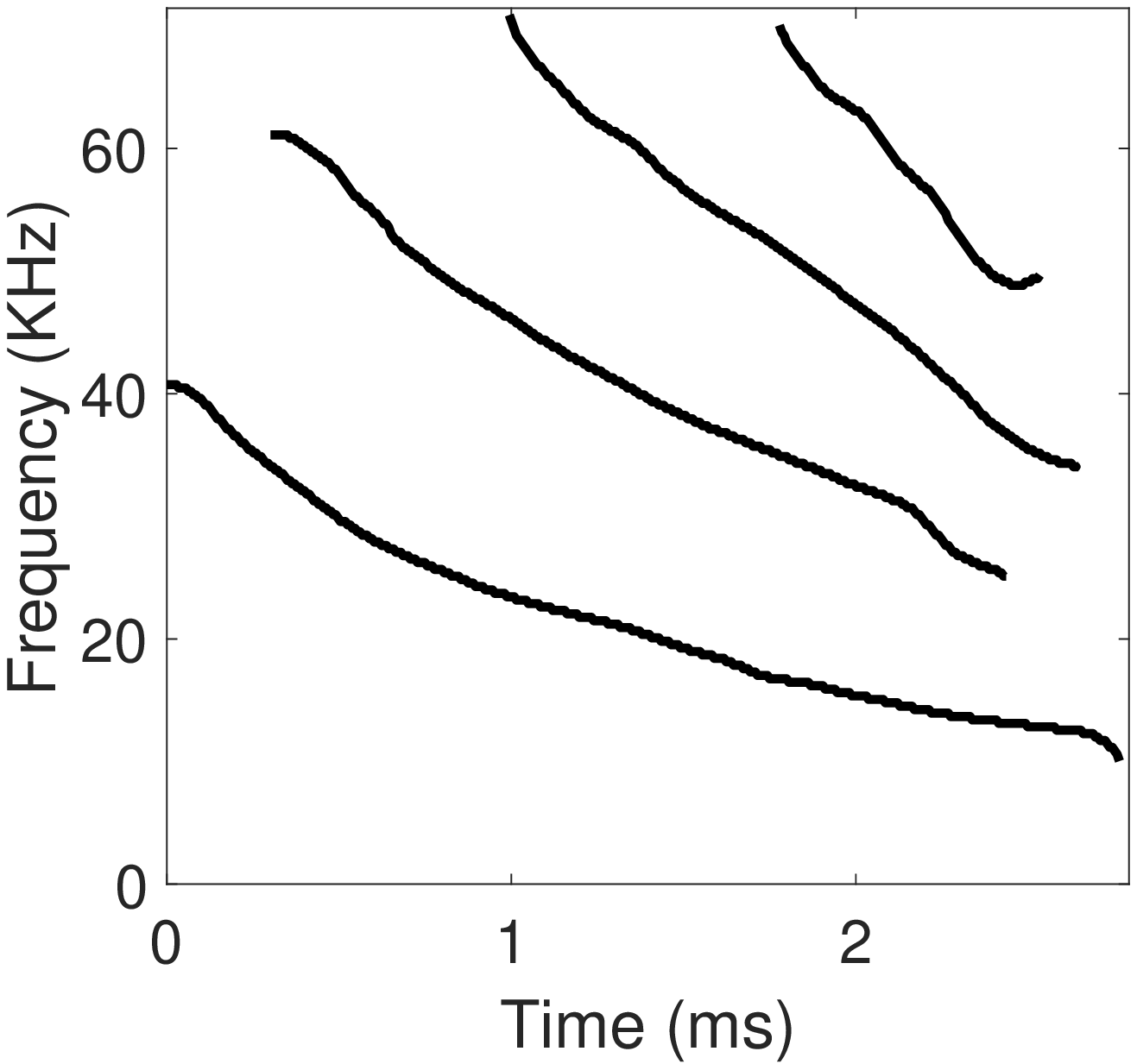}}
			\centerline{\fontsize{8.0pt}{\baselineskip}\selectfont (c)}
		\end{minipage}
		\hfill
		\begin{minipage}{0.48\linewidth}
			\centerline{\includegraphics[width=6.6cm]{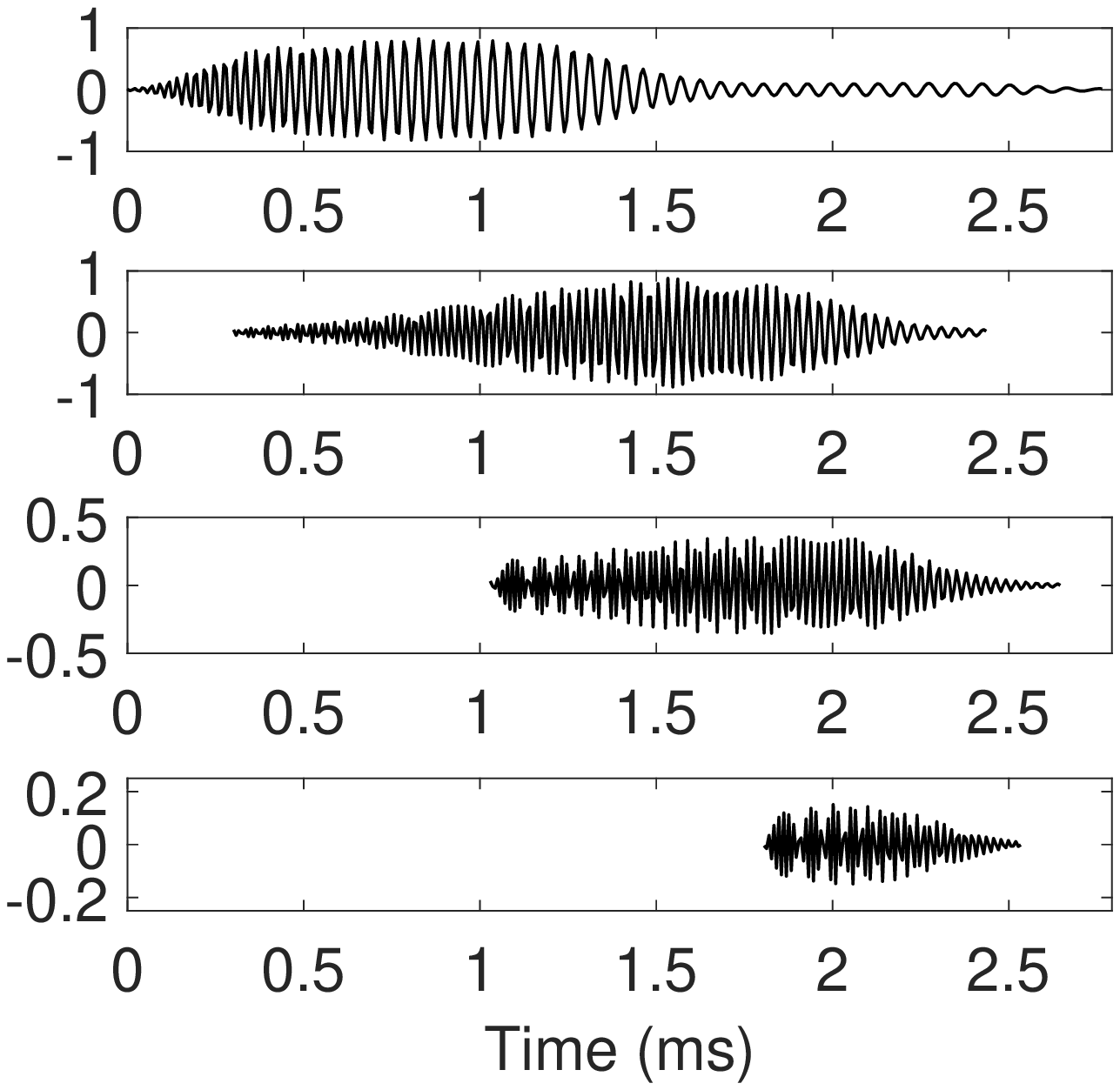}}
			\centerline{\fontsize{8.0pt}{\baselineskip}\selectfont (d)}
		\end{minipage}
		
		\caption {Bat signal and its estimation results: source data (a), spectrum (b), IF estimation (c) and waveform reconstruction (d).}
		\label{fig:bat_signal}
	\end{figure}

	Finally we test our method on a bat echolocation signal emitted by a large brown bat (Eptesicus Fuscus) in real world. 
	Fig.\ref{fig:bat_signal} (a) and (b) show the waveform and spectrum of the Bat echolocation signal, respectively.
	There are 400 samples with the sampling rate 143 KHz.
	The data can be downloaded from the website of DSP at Rich University \cite{bat}. The IF representation of this bat signal has been studied with the second-order FSST \cite{LCHJJ18} and the instantaneous frequency-embedded synchrosqueezing wavelet transform \cite{Jiang_Suter17} respectively. From the results in \cite{Jiang_Suter17}, we know the bat echolocation signal is made of four components, and all the components are approximated to linear chirp modes.
	
	\clearpage
	
	Fig.\ref{fig:bat_signal}(c) and (d) show the results of IF estimation and waveform reconstruction, respectively. Observe that the four components are decomposed clearly. Different from the existing ridge detection methods, which usually assumes the number of components is known, we use Algorithm 1 to extract each ridge by the local maxima without any prior knowledge. 
	
 The proposed method is suitable to process sub-signals with different supported time intervals,  such as the results in Fig.\ref{fig:music_data}(d) and Fig.\ref{fig:bat_signal}(d).
	The EMD methods cannot decompose both music signal and the bat signal above correctly.  
	Since the space limitation, we will not give these experiment results in this paper.		

\section{Conclusion}
In this paper, we introduce a direct signal separation method via extraction of local frequencies. 
We derive  more accurate component recovery formulae based on the linear chirp signal local approximation. A recovery scheme, together with a ridge detection method,  is also proposed to extract the signal components one by one, and the time-varying parameter is updated for each component.
The proposed method works like the EMD approach, can separate complicated multicomponent non-stationary signal adaptively and automatically without any prior knowledge. By the approximation of linear chirp modes and derivation of separation conditions, we show the proposed method is capable to separate components with closer IFs than the existing state-of-the-art methods. 
Moreover, for real signals, the proposed method can be implemented by FFT, which is suitable for engineering applications.      
The further study is to consider how to take advantage of the ASSO for separating multicomponent signals with crossings of instantaneous frequency curves.

\section*{Acknowledgment}
L. Li would like to thank Dr. Ningning Han and Prof. Hongbing Ji for helpful discussions. The authors would like to thank the anonymous reviewers who provide their valuable suggestions and comments which have improved the presentation of this paper.



\begin{thebibliography}{1}
\bibitem{Flandrin10} P. Flandrin  and P. Borgnat, ``Time-frequency energy distributions meet compressed sensing," {\it IEEE Transactions on Signal Processing}, vol. 58, no. 6, pp.2974-2982, June  2010.

\bibitem{AM_FM_Francesco07} F. Gianfelici, G. Biagetti, P. Crippa and C. Turchetti, ``Multicomponent AM-FM representations: An asymptotically exact approach," {\it IEEE Transactions on Audio, Speech and Language Processing}, vol. 15, no. 3, pp. 823--837,  Mar. 2007.

\bibitem{Huang98}  N.E. Huang, Z. Shen, S.R. Long, M.L. Wu, H.H. Shih, Q. Zheng, N.C. Yen, C.C. Tung,  and H.H. Liu, ``The empirical mode decomposition and Hilbert spectrum for nonlinear and nonstationary time series analysis,"  {\it Proc. Roy. Soc. London A}, vol. 454, no. 1971, pp. 903--995, Mar. 1998.

\bibitem{Chui_M15}  C. K. Chui and H. N. Mhaskar,  ``Signal decomposition and analysis via extraction of frequencies," {\it Appl. Comput. Harmon. Anal.}, vol. 40, no. 1, pp. 97--136, 2016.

\bibitem{Daub_Lu_Wu11} I. Daubechies, J. Lu, and H.-T. Wu, ``Synchrosqueezed wavelet transforms:
An empirical mode decomposition-like tool,"  {\it Appl. Computat. Harmon. Anal.}, vol. 30, no. 2, pp. 243--261, Mar. 2011.


\bibitem{Wu_thesis} H.-T. Wu, {\it Adaptive analysis of complex data sets,}   Ph.D. dissertation, Princeton Univ., Princeton, NJ, 2012.


\bibitem {De Prony} R. Prony, ``Essai experimental et analytique sur les Lois de la dilatabilite de fluides elastiques et sur celles de la force expansive de la varpeur de L'alkoal, a differentes temperatures," {\it J. l'Ecole Polytech.}, vol.1, no. 22, pp. 24 -- 76, 1795.


\bibitem{MUSIC} R.O. Schmidt, \lq\lq{}Multiple Emitter Location and Signal Parameter Estimation,"  IEEE Trans. Antennas and Propagation, vol. 34, no. 3,  pp. 276--280, Mar. 1986.

\bibitem{ESPRIT} R. Roy and T. Kailath, "ESPRIT-estimation of signal parameters via rotational invariance techniques,"  IEEE Trans.  Acoustics, Speech, and Signal Proc., vol. 37, no. 7, pp. 984--995, Jul. 1989.


\bibitem{Daub_Maes96} I. Daubechies and S. Maes,  ``A nonlinear squeezing of the continuous wavelet transform based on auditory nerve models,"  in A. Aldroubi, M. Unser  Eds. {\it Wavelets in Medicine and Biology}, CRC Press, 1996, pp. 527–-546.


\bibitem{Thakur_Wu11} G. Thakur and H.-T. Wu,  ``Synchrosqueezing-based recovery of instantaneous frequency from nonuniform samples," {\it SIAM J. on Mathematical Analysis}, vol. 43 No.5, 2078-2095, 2011.

\bibitem{Flandrin_Wu_etal_review13} F. Auger, P. Flandrin, Y. Lin, S.McLaughlin, S. Meignen, T. Oberlin, and H.-T. Wu, ``Time-frequency reassignment and synchrosqueezing: An overview,"  {\it IEEE Signal Process. Mag.}, vol. 30, no. 6, pp. 32--41, 2013.

\bibitem{Li_Liang12} C. Li and M. Liang, ``A generalized synchrosqueezing transform for enhancing signal time-frequency representation,"  {\it Signal Proc.}, vol. 92, no. 9, pp. 2264--2274, 2012.

\bibitem{Thakur_etal_Wu13} G. Thakur, E. Brevdo, N. Fu$\check{\rm c}$kar, and H.-T. Wu, ``The synchrosqueezing algorithm for time-varying spectral analysis: Robustness properties and new paleoclimate applications,"  {\it Signal Proc.}, vol. 93, no. 5, pp. 1079--1094, 2013.



\bibitem{MOM14} T. Oberlin, S. Meignen, and V. Perrier,  ``The Fourier-based synchrosqueezing transform,"  in {\it 2014 IEEE Int. Conf. Acoust., Speech, Signal Process. (ICASSP)},  May 2014, Florence, Italy, pp. 315--319.

\bibitem{Wang_etal14} S. Wang, X. Chen, G. Cai, B. Chen, X. Li, and Z. He, ``Matching demodulation transform and synchrosqueezing in time-frequency analysis," {\it IEEE Trans. Signal Proc.}, vol. 62, no. 1, pp. 69--84, 2014.


\bibitem{Chui_Walt15} C. K. Chui and M. D. van der Walt, ``Signal analysis via instantaneous frequency estimation of signal components,"  {\it Int'l  J  Geomath}, vol. 6, no. 1, pp. 1--42, Apr. 2015.


\bibitem{Yang15} H.Z. Yang, ``Synchrosqueezed wave packet transforms and diffeomorphism based spectral analysis for 1D general mode decompositions," {\it Appl Comput. Harmon. Anal.}, vol. 39, no.1,  pp. 33--66, 2015.



\bibitem{Chui_Lin_Wu15} C. K. Chui, Y.-T. Lin, and H.-T. Wu, ``Real-time dynamics acquisition from irregular samples– with application to anesthesia evaluation," {\it Anal. Appl.}, vol. 14, no. 4,  pp. 537--590,   Jul. 2016. 


\bibitem{MOM15} T. Oberlin, S. Meignen, and V. Perrier, ``Second-order synchrosqueezing transform or invertible reassignment? towards ideal time-frequency representations,"  {\it  IEEE Trans. Signal Proc.},
vol. 63, no. 5, p.1335--1344, Mar. 2015.


\bibitem{Jiang_Suter17} Q. Jiang and B.W.  Suter,  ``Instantaneous frequency estimation based on synchrosqueezing wavelet transform,'' {\it  Signal Proc.}, vol. 138,   pp.167--181, 2017.

\bibitem{Yang18} H.Z. Yang, ``Statistical analysis of synchrosqueezed transforms," {\it Appl. Comput. Harmon. Anal.}, vol. 45, no. 3, pp. 526--550, Nov. 2018.


\bibitem{WCSGTZ18} S.B. Wang, X.F. Chen, I.W. Selesnick, Y.J. Guo,  C.W. Tong and X.W. Zhang,
``Matching synchrosqueezing transform: A useful tool for characterizing signals with fast varying instantaneous frequency and application to machine fault diagnosis," {\it Mechanical Systems and Signal Proc.},
vol. 100, pp. 242--288, Feb. 2018.

\bibitem{LCHJJ18}  L. Li, H.Y. Cai, H.X. Han, Q.T. Jiang and H.B. Ji, ``Adaptive short-time Fourier transform and synchrosqueezing transform for non-stationary signal separation,\rq\rq{} {\it Signal Proc.}, vol. 166, Jan. 2020, Article 107231. 


\bibitem{LCJ18}  L. Li, H.Y. Cai, and Q.T. Jiang, ``Adaptive synchrosqueezing transform with a time-varying parameter for non-stationary signal separation,\rq\rq{} {\it Appl. Comput. Harmon. Anal.}, vol. 49, no. 3,, pp. 1075--1106, Nov. 2020. 

\bibitem{CJLS21} H.Y. Cai, Q.T. Jiang, L. Li and B.W.  Suter,  ``Analysis of adaptive short-time Fourier transform-based synchrosqueezing transform,\rq\rq{} {\it Analysis and Applications}, vol. 19, no. 1, pp. 71--105, 2021. 

\bibitem{LJL21} J. Lu, Q.T. Jiang and L. Li, ``Analysis of adaptive synchrosqueezing transform with a time-varying parameter,\rq\rq{}  {\it Adv. Comput. Math.}, vol. 46, Article 72, 2020.


\bibitem{Flandrin04} P. Flandrin, G. Rilling, and P. Goncalves, ``Empirical mode decomposition as a filter bank," {\it IEEE Signal Proc. Letters}, vol. 11,  pp. 112--114, Feb. 2004.

\bibitem{Xu06} Y. Xu, B. Liu, J. Liu, and S. Riemenschneider, ``Two-dimensional empirical mode decomposition by finite elements," {\it Proc. Roy. Soc. London A}, vol. 462, no. 2074, pp. 3081--3096, Oct. 2006.

\bibitem{Rilling08} G. Rilling and P. Flandrin, ``One or two frequencies? The empirical mode decomposition answers," {\it IEEE Trans. Signal Proc.}, vol. 56, pp. 85--95, Jan. 2008.

\bibitem{Wu_Huang09} Z. Wu and N.E. Huang, ``Ensemble empirical mode decomposition: A noise-assisted data analysis method,''  {\it Adv. Adapt. Data Anal.}, vol. 1, no. 1, pp. 1--41,  Jan. 2009.

\bibitem{Li_Ji09}L. Li and H. Ji, ``Signal feature extraction based on improved EMD method," {\it Measurement}, vol. 42, pp. 796--803, June 2009.

\bibitem{HM_Zhou09} L. Lin, Y. Wang, and H.M. Zhou, ``Iterative filtering as an alternative algorithm for empirical mode decomposition,''  {\it Adv. Adapt. Data Anal.}, vol. 1, no. 4, pp. 543--560, Oct. 2009.

 \bibitem{Y_Wang12} Y. Wang, G.-W. Wei and S.Y. Yang , ``Iterative filtering decomposition based on local spectral evolution kernel, {\it J. Scientific Computing}, vol. 50, no. 3, pp. 629--664, Mar. 2012.


\bibitem{HM_Zhou16} A. Cicone, J.F. Liu, and H.M.  Zhou, ``Adaptive local iterative filtering for signal decomposition and instantaneous frequency analysis,'' {\it Appl. Comput. Harmon. Anal.}, vol. 41, no. 2, pp. 384--411, Sep. 2016.

\bibitem{Li_Jiang19}L. Li and H. Cai, Q. Jiang and H. Ji, ``An empirical signal separation algorithm for multicomponent signals based on linear time-frequency analysis," {\it Mechanical Systems and Signal Processing}, vol. 121, no. 4, pp. 791--809,  Apr. 2019.


 \bibitem{Cicone20} A. Cicone. ``Iterative Filtering as a direct method for the decomposition of nonstationary signals,\rq\rq{} {\it Numerical Algorithms}, vol. 373, 112248, 2020. 


\bibitem{van20} M.D. van der Walt, ``Empirical mode decomposition with shape-preserving spline interpolation,\rq\rq {\it Results in Applied Mathematics}, vol. 5, 100086, Feb. 2020. 


\bibitem{HM_Zhou20} A. Cicone and H.M. Zhou, ``Numerical analysis for iterative filtering with new efficient implementations based on FFT,\rq\rq{} {\it Numer. Math.}, vol. 147, pp. 1--28, 2021.


\bibitem{Bhattacharyya18}	
A. Bhattacharyya, L. Singh and R. B. Pachori, ``Fourier–Bessel series expansion based empirical wavelet transform for analysis of non-stationary signals,\rq\rq{} {\it  Digital Signal Processing}, vol. 78, pp. 185-196, Jul. 2018.
	
\bibitem{Upadhyay20}
A. Upadhyay, M. Sharma, R. B. Pachori and R. Sharma,  ``A nonparametric approach for multicomponent AM–FM signal analysis,\rq\rq{} {\it Circuits, Systems, and Signal Processing}, vol. 39, no. 12, pp. 6316--6357, 2020.

\bibitem{Pachori10}
R. B. Pachori and  P. Sircar,  ``Analysis of multicomponent AM-FM signals using FB-DESA method,\rq\rq{} {\it Digital Signal Processing}, vol. 20, no. 1, pp. 42--62, Jan. 2010.

\bibitem{Jain15}
P. Jain and R. B. Pachori,  ``An iterative approach for decomposition of multi-component non-stationary signals based on eigenvalue decomposition of the Hankel matrix,\rq\rq{} {\it Journal of the Franklin Institute}, vol. 352, no. 10, pp. 4017--4044, Oct. 2015.

\bibitem{Sharma18}
R. R. Sharma and R. B. Pachori, ``Time-frequency representation using IEVDHM–HT with application to classification of epileptic EEG signals,\rq\rq{} {\it IET Science, Measurement \& Technology}, vol. 12, no. 1, pp. 72--82, 2018


\bibitem{Stankovic01a} L. Stankovi${\acute {\rm c}}$, M. Dakovi${\acute {\rm c}}$  and V. Ivanovi${\acute {\rm c}}$, ``Performance of spectrogram
as IF estimator,\rq\rq{} {\it Electronics Letters}, vol. 37, no. 12, pp. 797--799, 2001.

\bibitem{Stankovic08} E. Sejdi${\acute {\rm c}}$,  I. Djurovi${\acute {\rm c}}$,  and L. Stankovi${\acute {\rm c}}$, ``Quantitative performance analysis of scalogram as instantaneous frequency estimator,\rq\rq{} {\it IEEE Trans. Signal Proc.},  vol. 56, no. 8, pp. 3837--3845, Aug. 2008. 

\bibitem{IMS15} D. Iatsenko, P.-V. E. McClintock and A. Stefanovska, ``Linear and synchrosqueezed time-frequency representations revisited: Overview, standards of use, resolution, reconstruction, concentration, and algorithms," {\it Digital Signal Proc.}, vol. 42, pp. 1--26, Jul. 2015.

\bibitem{table_book} H. Bateman, {\it Tables of Integral Transforms}, Vol. I, 1954, McGraw-Hill, New York. 

 \bibitem{CJLL20} C.K. Chui, Q. Jiang, L. Li and J. Lu, ``Analysis of  an  adaptive short-time Fourier transform-based multi-component signal separation method derived from  linear chirp local approximation,\rq\rq{} {\it J. Comput. Appl. Math.}, vol. 396, 113607, Nov. 2021.

\bibitem{LM1} N. Laurent and S. Meignen, ``A novel time-frequency technique for mode
retrieval based on linear chirp approximation,\rq\rq{} {\it IEEE Signal Proc. Letters}, vol. 27, 
pp. 935--939, May 2020. 

\bibitem{LM2} N. Laurent and S. Meignen, ``A novel approach for ridge detection and
mode retrieval of multicomponent signals based on STFT,\rq\rq{} preprint, Sep. 2020. arXiv:2009.13123v1.  



\bibitem{Leon_Cohen}  L. Cohen, {\it Time-frequency Analysis}, Prentice Hall, New Jersey, 1995. 

\bibitem{Baraniuk01} R. Baraniuk, P. Flandrin, A. Janssen, and O. Michel,  ``Measuring time-frequency information content using the R${\rm \acute e}$nyi entropies," {\it  IEEE Trans. Inform. Theory}, vol. 47, no. 4, pp. 1391--1409, 2001.



\bibitem{Stankovic01}L. Stankovic, ``A measure of some time-frequency distributions concentration,\rq\rq{} {\it Signal Proc.}, vol. 81, no. 3, pp. 621--631, 2001.


\bibitem{Wu17} Y.-L. Sheu, L.-Y. Hsu, P.-T. Chou, and H.-T. Wu, ``Entropy-based time-varying window width selection for nonlinear-type time-frequency analysis," {\it Int'l J. Data Sci. Anal.}, vol. 3,  pp. 231--245, 2017.

\bibitem{SP18} V. Sharma and A. Parey, ``Performance evaluation of decomposition methods to diagnose leakage in a reciprocating compressor under limited speed variation,\rq\rq{} {\it Mechanical Systems and Signal Proc.}, vol. 125, pp. 275--287, Jun. 2019.

\bibitem{Li_Ji06} L. Li and H. Ji, ``Radar targets detection in formation based on time-varying AR model,''  in {\it  2006 CIE International Conference on Radar (ICA)}, October 2006, Shanghai, China.

\bibitem{bat} https://web.archive.org/web/20160403234536/http://dsp.rice.edu/software/bat-echolocation-chirp  . 

\end{thebibliography}
\end{document}